\documentclass[twocolumn,showpacs,preprintnumbers,amsmath,amssymb,prb,floatfix]{revtex4}

\usepackage{graphicx,tabularx}
\usepackage{amssymb}
\usepackage{dcolumn}
\usepackage{color}% to create grayscales
\usepackage[mathcal]{euscript}
\def\be{$\beta_{1ud}$}
\def\btu{$\beta_{2u}$}
\def\btd{$\beta_{2d}$}
\def\bd{$\beta_{3ud}$}
\def\bvu{$\beta_{4u}$}
\def\bvd{$\beta_{4d}$}
\def\bf{$\beta_{5ud}$}
\def\bsu{$\beta_{6u}$}
\def\bsd{$\beta_{6d}$}
\def\bz{$\beta_{7ud}$}

\def\ge{$\gamma_{1}$}
\def\gt{$\gamma_{2}$}
\def\gd{$\gamma_{3}$}

\def\cge{$c(4\times2)$}
\def\brec{$b(2\times1)$-reconstruction\ }

\def\crec{$c(4\times2)$-reconstruction\ }

\def\bsurec{$\beta_{6u}$-reconstruction\ }
\def\bsugeom{$\beta_{6u}$-geometry\ }

\definecolor{gray0}{gray}{0.0}%black
\definecolor{gray64}{gray}{0.25}
\definecolor{gray128}{gray}{0.5}
\definecolor{gray192}{gray}{0.75}
\definecolor{gray255}{gray}{1.0}%white

\begin{document}
\title{DFT study of Pt-induced Ge(001) reconstructions}
\author{Danny E. P. Vanpoucke}
\affiliation{Computational Materials Science, Faculty of Science and Technology and MESA+ Institute for Nanotechnology, University of Twente, P.O. Box 217, 7500 AE Enschede, The Netherlands}
\author{Geert Brocks}
\affiliation{Computational Materials Science, Faculty of Science and Technology and MESA+ Institute for Nanotechnology, University of Twente, P.O. Box 217, 7500 AE Enschede, The Netherlands}

\date{\today}
\begin{abstract}
Pt deposited on a Ge(001) surface spontaneously forms nanowire arrays.
These nanowires are thermodynamically stable and can be hundreds of
atoms long. The nanowires only occur on a reconstructed Pt-Ge-surface
where they fill the troughs between the dimer rows on the surface. This
unique connection between the nanowires and the underlying substrate
make a thorough understanding of the latter necessary for understanding
the growth of the nanowires. In this paper we study possible
surface reconstructions containing $0.25$ and $0.5$ of a monolayer of Pt. Comparison of calculated STM images to experimental STM images of the surface
reconstruction reveal that the Pt atoms are located in the top layer,
creating a structure with rows of alternating Pt-Ge and Ge-Ge dimers in a $c(4\times2)$ arrangement. Our results also show that Pt atoms in the second
or third layer can not be responsible for the experimentally observed
STM images.
\end{abstract}

 %--> wtf???
%----------------------------------------------------------------
\maketitle
%-----------------------------------------------------------------
%------------------------------------------------------------------------------------------
%------------------------------Introduction------------------------------------------------
%------------------------------------------------------------------------------------------
% link to pt nanowires? lack of theory and experimental info?
\section{Introduction}
Metal/semiconductor interface interactions continue to attract interest
in both fundamental and applied research, and are of vital importance
for the semiconductor and microelectronics industry. With
miniaturization pushing into the nanoscale regime, new ways of constructing nanoscale devices are investigated. Bottom-up approaches such as atomic size reconstructions, self assembly and metal nanowires (NWs) are investigated as alternatives for the usual top-down techniques, such as etching and lithography.\cite{Barth:nat05}
Generally, the focus lies on metal/Si interfaces because of the importance of Si in the semiconductor industry. Of these, the Au/Si(001) interface, due to the high conductivity of Au, is one of the most well studied interfaces.\\
\indent For this system, it has been shown that different types of
reconstructions appear depending on the Au coverage and the annealing
temperature.\cite{LinXF:prb93,Kageshima:ss01,Shimakura:ss98} The
metal/Ge interface is less well studied, notwithstanding its importance in the
development of radiation detector systems and high speed electronic
devices. Recent experimental studies of the Pt/Ge(001) interface, done
at the submonolayer Pt coverage regime, have shown the formation of
self-organized Pt NW arrays after high temperature annealing.\cite{Gurlu:apl03,Oncel:prl05,Schafer:prb06} It has been suggested that this reconstruction was due to a strengthened interaction of the 5$d$
metals with the Ge surface: relativistic mass effects which contract
the $s$ shell, reducing its energy and thus increase the $s$ occupancy
at the expense of the $d$ electrons. This partially depletes the
antibonding $d$ states, strengthening the $d$ bonds.\cite{Smit:prl01}
The same mechanism was thought to be responsible for the one
dimensional chain structures of Au on Ge.\cite{Wang:prb04,Wang:ss05}
There certainly are similarities between the growth of Au and Pt chains
on Ge(001), such as the appearance of dimer vacancies before
the chains appear. Yet, there are also important differences between
the two systems. The Pt NWs only form after annealing at over 1000 K,
while the Au chains appear after deposition at 675 K and disappear into
the bulk Ge when annealed at 1000 K.\cite{Wang:ss05,Gurlu:apl03}
Moreover, while the Pt NWs have a thickness of a single atom, the Au
chains are dimer rows containing Au-Au or Au-Ge
dimers.\cite{Wang:prb04} The Pt NWs are thermodynamically stable and
their length is only limited by the size of the underlying terrace.
This so-called $\beta$-terrace is a previously unknown Pt-Ge-surface reconstruction of which, to date, only a tentative model proposed by G\"url\"u
\textit{et al.} exists.\cite{Gurlu:apl03} The unique connection
between this terrace and the NWs makes a thorough understanding of the
reconstruction desirable before one can study and understand the NW
arrays and the physical phenomena associated with
them.\cite{vanpoucke:prb2008R,stekolnikov:prl08}\\
\indent In this work, we present a geometry for the $\beta$-terrace based on the comparison of the simulated scanning tunneling microscope (STM) images
(henceforth referred to as pseudo-STM images) to experimental STM
images. The current experimental information on the
$\beta$-terraces is very limited. At the moment of writing only
STM images are available. A preliminary account of this work appeared in Ref.~\onlinecite{vanpoucke:prb2008R}.\\
\begin{figure}[!bp]
\begin{center}
  \includegraphics[width=8.50cm,keepaspectratio]{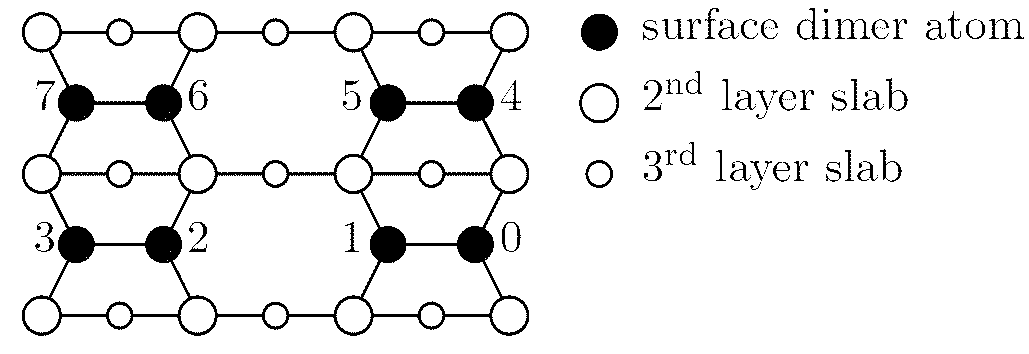}\\
\end{center}
\caption{Top view of the Ge(001) surface with symmetric $(2\times1)$
reconstruction. 0--7 are indexes for the surface dimer atoms and are
used in the nomenclature of the geometries (see text).}
\label{fig:bnomen}
\end{figure}
\indent The structure of this paper is as follows: In Sec.~\ref{sc:theormeth} we describe the theoretical method and introduce the notation used for the different geometries. In Sec.~\ref{sc:theorres} we present our theoretical results starting with a comparison of the formation energies of the different geometries
followed by a general review of the pseudo-STM images. We identify one particular geometry as the structure of the $\beta$-terrace, which is discussed in more detail. As a last point of Sec.~\ref{sc:theorres}, we take a look at some geometries containing Pt atoms in the second and third layer and show that these can not be responsible for the $\beta$-terrace STM pictures, but they could be responsible for the indented dimer and two-dimer vacancy
images seen in the STM pictures of a different Pt induced Ge(001) reconstruction, the so-called $\alpha$-terrace.\cite{Gurlu:apl03} Finally, in Sec.~\ref{sc:conclusions} the conclusions are given.
%------------------------------------------------------------------------------------------
%------------------------------Theoretical_method------------------------------------------
%------------------------------------------------------------------------------------------
\section{Theoretical method}\label{sc:theormeth}
%Theoretical setup and nomenclature
\subsection{Setup}\label{ssc:theormeth_setup}
%--> setup of calculation: tech data
Density functional theory (DFT) calculations are carried out using the projector augmented waves (PAW) method and the local density approximation (LDA) functional with the exchange and correlation function calculated by Ceperley and Alder (CA), as implemented in the VASP program.\cite{Blochl:prb94,Kresse:prb99,CA:prl1980,Kresse:prb93,Kresse:prb96} A plane wave basis set with kinetic energy cutoff of $287$ eV is applied and the semicore
$d$ shell of Ge is included in the valence shell for better accuracy.
The surface is modeled by periodically repeated slabs of $12$ layers of
Ge atoms. Extensive convergence tests, showed no advantage in time/accuracy for a H-passivated Ge slab over a symmetric Ge slab. Based on the advantages of the higher symmetry of the symmetric slab, we choose to use the later. The slabs are then mirrored in the $z$-direction with reconstructions on both surfaces, in which specific Ge atoms are replaced by Pt atoms. A vacuum region of $\sim15.5$ \AA\ is used to separate the slabs along the $z$ axis. The Brillouin zone (BZ) of the $(2\times4)$ surface unit cell is sampled using a $6\times3$ Monkhorst-Pack special $k$-point mesh.\cite{Monkhorst:prb76} To optimize the geometry of the slabs the conjugate gradient method is used. The positions of the Ge atoms in the two center layers are fixed as to represent the bulk phase of the system.
\subsection{Ge(001)-surface and nomenclature}\label{ssc:theormeth_nomen}
%% nomenclature
%%   Ge(001) reconstructions
%%   int + ud indexes
To model the $\beta$-terrace we first model the clean
Ge(001)-surface. Both the buckled $b(2\times1)$ and $c(4\times2)$
reconstructions using a $(4\times2)$ supercell and
the symmetric $(2\times1)$ reconstruction were generated using a
$(2\times1)$ supercell. After relaxation, the
$b(2\times1)$ and the $c(4\times2)$ reconstruction were found to be
$0.302$ and $0.374$ eV per dimer lower in energy than the symmetric
$(2\times1)$ dimer reconstruction. A buckling angle of $\sim
19.5^{\circ}$ was found for both asymmetric reconstructions, and a
dimer length of $2.45$ \AA\ for the $b(2\times1)$ reconstruction and of
$2.51$ \AA\ for the $c(4\times2)$ reconstruction was calculated, all in
accordance with previous theoretical and experimental
results.\cite{Needels:prl87,Needels:prb88,Jenkins:jp96,Gay:prb99}\\
\indent Based on experimental evidence, \cite {Gurlu:apl03,Oncel:prl05} we
assume that the $\beta$-terrace contains $0.25$ monolayer (ML) of Pt. Furthermore, we assume that these Pt-atoms are located in the top layer of the Ge slab. We will show in Sec.~\ref{ssc:theorres_othergeom} that
Pt in the second or third layer can not be responsible for the
experimentally obtained STM images.\\
\begin{table}[!t]
\center{\textbf{Formation energies and buckling angles.}}
\begin{ruledtabular}
\begin{tabular}{l|ccc }
& $E_f$ & Pt-Ge buckling & Ge-Ge buckling \\
& (meV) & (degrees) & (degrees) \\
\hline
$b(2 \times 1)$ & $0$ & - & $19.42$\\
$c(4 \times 2)$ & $-73$ & - & $19.67$\\
\hline
\be & $589$ & $20.57$ & $19.07$ ($18.25$) \\
\btu & $-16$ & $-4.42$ & $21.03$ \\
\btd & $94$ & $-0.15$ & $18.23$ \\
\bd & $34$ & $-4.43$(u) & $21.85$(u) \\
& & $-1.82$(d) & $18.13$(d) \\
\bvu & $-61$ & $1.92$ & $18.85$ \\
\bvd & $-25$ & $0.50$ & $18.85$ \\
\bf & $83$ & $-5.04$(u) & $19.06$(u) \\
& & $7.38$(d) & $18.43$(d)\\
\bsu & $-25$ & $-4.25$ & $20.82$ \\
\bsd & $82$ & $-0.88$ & $18.13$ \\
\bz & $29$ & $-4.83$(u) & $21.43$(u) \\
& & $-1.30$(d) & $18.06$(d)\\
\hline
$\gamma_1$ ($2\times1$)  & $-47$ & $1.95$ & - \\
$\gamma_2$ ($2\times2$)  &  $80$ & $6.13$ , $-6.46$ & - \\
$\gamma_3$ ($4\times2$)  &  $89$ & $5.75$ , $-6.11$ & - \\
\end{tabular}
\end{ruledtabular}
\caption{The formation energy and dimer buckling angles of the $\beta$ and $\gamma$-surfaces compared to Ge(001) surface reconstructions. For the \be\ geometry the Pt-Ge dimer buckling refers to
the buckling of the Pt-Pt dimer and the value between brackets for the
Ge-Ge buckling refers to the buckling of the Ge-Ge dimers located in
the same dimer row as the Pt-Pt dimers. The notations (u) and (d) in
the column of the Pt-Ge buckling refer to the position of the Pt atom
in the dimer, while in the Ge-Ge buckling column they indicate if the
Ge dimer is in a dimer row with the Pt atom in the up or down position
(except for \bf\ where it refers to the location of the Pt atom in a
Pt-Ge dimer in the adjacent dimer row). For $\gamma_2$ and $\gamma_3$
the buckling angles of both dimers of a single dimer row are
given.}\label{table:geom_energ}
\end{table}
\indent To model the $\beta$-terrace, all possible non-isomorphic geometries are examined. We generate these geometries by substitution of two surface Ge atoms, of the $b(2\times1)$ surface reconstruction in the
$(4\times2)$ supercell, by Pt atoms.
Ignoring the buckling of the Ge dimers for a moment, this leads to
seven non-isomorphic geometries. The first Pt atom replaces the Ge
surface atom in position $0$, while the second Pt atom replaces one of
the other seven surface Ge atoms (\textit{cf.}\ Fig.~\ref{fig:bnomen}). We will refer to these geometries as $\beta_1$ to $\beta_7$. If the buckling of
the Ge dimers in the $b(2\times1)$ reconstruction is taken into
account, some of the symmetry is lost. Let us assume for a moment that
the atoms of the dimers in Fig.~\ref{fig:bnomen} with an even index are
the down dimer atoms and those with an odd index are the up dimer
atoms. In what follows we will keep referring to those
positions in this manner. We now notice that in some geometries both
Pt atoms are located at an up (down)-dimer atom position and in other
cases one Pt atom is located in an up dimer atom position while the
other Pt atom is located in a down dimer atom position.\cite{fn:dimernomen}
While for the latter all
combinations that are isomorphic stay isomorphic when
tilting is taken into account this is not the case in the former
situation. Take for example the $\beta_2$ geometry with its Pt atoms in
positions $0$ and $2$. An isomorphic geometry in the symmetric
reconstruction is one with Pt atoms in positions $1$ and $3$. In the
$b(2\times1)$ geometry, as described above, the first $\beta_2$
geometry has both Pt atoms at down dimer atom sites, while the second
has them at up dimer atom sites. To distinguish these two different
geometries, indices $u$ and $d$ are used to indicate the up or down
location of the Pt atoms. So the first $\beta_2$ geometry is called
\btd\ and the second one \btu.\cite{fn:udcomm}
For geometries with one atom at an up and one at a down dimer atom site
both the $u$ and $d$-index are used to indicate just this
(\textit{e.g.}\ $\beta_3\rightarrow\beta_{3ud}$). This results into
ten non-isomorphic geometries to start from.\\
\indent To complete the picture of this system we also included
possible geometries containing $0.5$ ML of Pt in the top layer. Such a
surface then consists entirely of Pt-Ge dimers. Geometries
containing pure Pt and pure Ge dimers were not considered based on our
results for the $\beta$-geometries.\\
\indent The three possible geometries considered are a $2\times1$
reconstruction ($\gamma_1$) with all Pt atoms at the even (or odd)
positions, a $2\times2$ reconstruction ($\gamma_2$) with the Pt atoms
at positions $0$, $2$, $5$, and $7$, and a $4\times2$ reconstruction
($\gamma_3$) with the Pt atoms at positions $0$, $3$, $5$, and
$6$.
All these geometries were optimized using the conjugate gradient method
while keeping the positions of the Ge atoms in the two center layers
fixed.
%------------------------------------------------------------------------------------------
%------------------------------theoretical_results-----------------------------------------
%------------------------------------------------------------------------------------------
\section{Theoretical results}\label{sc:theorres}
%Results
\subsection{Geometry and formation energy}\label{ssc:theorres_formen}
\subsubsection{Stability of the $\beta$-geometries}
%% formation energy and stuff
The stability of the different geometries was investigated by
comparison of their formation energies to the formation energy of the
Ge $b(2\times1)$ surface reconstruction. The formation energy of a
certain geometry, $\beta_{\mathrm{x}}$, can be calculated using the
expression:\cite{fn:altformenerg}
\begin{equation}
E_f=[(E_{\beta_{\mathrm{x}}}+N_{Pt}E_{Ge})-(E_{b(2\times1)}+N_{Pt}E_{Pt})]/N_{Pt-dimers} , \label{eq:1_Formation_energy}
\end{equation}
with $E_{\beta_{\mathrm{x}}}$ the total energy of the
$\beta_{\mathrm{x}}$-geometry, $N_{Pt}$ the number of Pt atoms in the
system, $E_{Ge}$ and $E_{Pt}$ the bulk energy of a Ge and a Pt atom,
$E_{b(2\times1)}$ the total energy of the Ge $b(2\times1)$ surface
reconstruction and $N_{Pt-dimers}$ the number of surface dimers
containing Pt atoms. Negative values for $E_f$ indicate an increase in
stability caused by replacing a Ge atom by a Pt atom.\\
\indent Table~\ref{table:geom_energ} clearly shows that formation energy of the \be-surface is much higher than that of the other structures. This is
the only structure containing a Pt-Pt dimer; all the other surface
models contain mixed Pt-Ge dimers. It shows that replacing a Ge atom in
a Pt-Ge dimer with a Pt atom will require about $0.5$ to $0.6$ eV. This
allows us to rule out the formation of Pt-Pt dimers in the $\beta$
terrace. Table~\ref{table:geom_energ} also shows a preference for
geometries which contain Pt-Ge dimer rows with the Pt atom in the $u$p
position. The entire set of formation energies, with the exception of \be,
is spread over a relatively small energy interval of $\sim150$ meV,
which is of the same order of magnitude as the thermal energy
corresponding to the annealing temperature. In the experiment, the germanium samples on which the Pt is deposited need to be annealed at a temperature of $>1000$ K ($\approx86$ meV) before the $\beta$-terrace (and the nanowires) appear. \cite{Gurlu:apl03} This makes it impossible to exclude any structures based solely on energetics, with the exception of the $\beta_{1ud}$-structure.\\
\indent All formation energies are given per Pt containing dimer. This
means that the formation energy is the average energy to form a Pt-Ge dimer in the
$\beta_2$--$\beta_7$ structures.\cite{fn:altformenerg}
In the structures indicated with the \textit{ud}-subscript the Pt-Ge
dimers are not equivalent. To make a distinction between these two types of dimers we will refer to a Pt-Ge dimer with the Pt atom in the up position as a Pt-up dimer, and similar for a Pt-down dimer. The \bd-structure contains a Pt-up and a Pt-down dimer, while the \btu- and \btd-structures contain either only Pt-up or Pt-down dimers.\\
\indent If we take the average of the formation energy of a Pt-up and a Pt-down dimer from the \btu- and \btd-structures we get $39$ meV, comparable to the formation energy of an average Pt-Ge dimer in the \bd-structure. Comparison of the \bz-structure with the \bsu- and \bsd-structures gives a similar result, $28.5$ meV for the average of the \bsu- and \bsd-structure compared to $29$ meV for the average \bz\ Pt-Ge dimer. The pattern of Pt-Ge and Ge-Ge dimers in these two sets of three geometries is the same, indicating that the energy due to interactions between neighboring dimer rows is small compared to the
formation energy. In contrast, the formation energy of the Pt-Ge dimer
depends strongly on the surrounding dimers within the dimer row.\\
\indent The third triplet of similarly patterned structures (\bf, \bvu\
and \bvd) shows this last point even more clearly. In this case the average of the formation energy of a Pt-Ge dimer of the \bvu- and \bvd-structure is
$-43$ meV, much less than the average formation energy for a Pt-Ge
dimer in the \bf-structure. Unlike the previous examples, the Pt-Ge
dimers of the \bf-structure have a different surrounding along the
dimer row than those in the $\beta_4$-structures: antiparallel instead
of parallel Pt-Ge dimers. This shows that the interaction between
neighboring dimers in a single dimer row is strong, and provides a
large contribution to the formation energy of a Pt-Ge dimer. It also
shows that rotation of a Pt-Ge dimer in the \bf-surface cell would give
an energy gain of roughly $0.2$--$0.3$ eV per surface unit cell.\\
%$216$ to $288$ meV per surface cell.\\
\indent Based on such comparisons we can make an estimate of the inter and intra dimer row energy contributions to the formation energy of a Pt-Ge dimer, $E_f$. To estimate the interaction energy between neighboring dimer rows, we compare structures with different inter dimer row neighbors for the Pt-Ge dimers. Comparison of the $\beta_2$- to the
$\beta_6$-structures, the \bd- to the \bz-structure, the $\beta_4$- to
the \ge-structure, the \gt- to the \gd-structure, and \bf- to the \gt- and \gd-structures, shows that this interaction energy is of the order of about $5$--$10$ meV per Pt-Ge dimer.\\
\indent A similar estimate can be made for the interaction energy
along a dimer row by comparing structures with different neighbors in a single dimer row for a Pt-Ge dimer. Comparing the $\beta_4$- to the $\beta_6$- and the \bf-structures, and comparing the \ge- to the \gt- and \gd-structures, gives an interaction energy in the range of $50$--$150$ meV per Pt-Ge
dimer, about one order of magnitude larger than the inter-dimer-row
interaction energy. It also shows that the proximity of the second Pt
atom has a large influence, both for better (\bvu) and for worse (\bf),
on the formation energy. In case of the \bvu-reconstruction all Pt atoms are located at the same side of the dimer row while for the \bf-reconstruction the Ge and Pt atoms of the dimer rows have a checkerboard pattern with the resulting surface strain deteriorating the formation energy.\\
%% geometries (dimerlength & angles)
\indent Furthermore, we found that the length of the dimers was not
influenced by the different surface structures, with the sole exception of \be, which is the only surface structure containing a Pt-Pt dimer. The lengths
of the Ge-Ge dimers were found to be $2.45\pm0.01$ \AA\ ($2.34\pm0.03$ \AA\ for \be) and for the Pt-Ge dimers $2.35$ \AA\ ($2.58$ \AA\ for the
Pt-Pt dimer in \be). The buckling angles on the other hand do
show a dependence on the location of nearby Pt atoms. Noteworthy is the
fact that the Pt-Ge dimers show little to no buckling and the little
buckling effect they show depends on the location of the Pt atom: if
the Pt atom is located in the up position then the buckling angle is
about $4$--$5\mathrm{^{\circ}}$, while it is less than
$2\mathrm{^{\circ}}$ if the Pt atom is located in the down position.
The sign used for the buckling angles indicates wether dimers are
buckled in the same (same sign) or opposite (opposite sign) direction of the Ge surface dimers. In all cases, except \be\ and $\beta_4$, the Pt-Ge dimers are buckled in the opposite direction of the Ge-Ge dimers thus creating a
$(2\times2)$ or $(4\times2)$ surface reconstruction. These
reconstructions also modify the Ge positions in the second and third
layer in a similar way as found for the Ge $c(4\times2)$ surface
reconstruction, which will be discussed in more detail in Sec.~\ref{ssc:theorres_beta6u}.\cite{Needels:prl87,Gay:prb99} Table~\ref{table:geom_energ} also shows that the Ge-Ge dimers are slightly
influenced by the Pt-Ge dimers in the same dimer row, increasing the
buckling angle to about $21$--$22\mathrm{^{\circ}}$ for Pt-Ge
dimers with the Pt atom in the up position and decreasing it to about
$18\mathrm{^{\circ}}$
if the Pt atom in the Pt-Ge dimer is in the down position.\\
\indent Another difference between Pt-Ge dimers with the Pt atom in the
up or down position becomes clear when the geometry of the dimers of a dimer
row containing both Pt-Ge and Ge-Ge dimers is investigated. The Pt-Ge dimer shifts
%(average of $t_3$ and $t_4$ in Fig.~\ref{fig:bondangledef}
%and Table~\ref{table:bonds})
orthogonal to the dimer row. If the Pt atom is
located in the up position this shift is very small ($\sim0.03$ \AA)
and almost entirely due to the Pt atom moving in the direction of the
up position. On the other hand, if the Pt atom is located in the down position, the shift is in the down direction and both the Ge and the Pt atom of the Pt-Ge dimer are shifted over a considerably larger distance of $\sim0.4$ \AA.\\
\indent Comparison of the average $z$-position of the dimers shows the
Pt-Ge dimers to be $0.01$ to $0.14$ \AA\ lower than the Ge-Ge
dimers for each of the considered structures.
%%%% Comparing the
%%%% positions of the centers of the dimers showed that in each case
%%%% the center of the Pt-Ge dimer was located $0.01$ to $0.14$ \AA\
%%%% more inward the slab than the center of the Ge dimers.
All the geometrical information of the Pt-Ge dimers combined with
the formation energies given in Table~\ref{table:geom_energ} shows
a preference for Pt atoms to be surrounded by Ge atoms which stick further out of the surface than these Pt atoms. This agrees with the preference for Ge-terminated surfaces in Pt/Ge alloys. \cite{Niranjan:prb07}
%------------------------------------------------------------------------------------------
%------------------------------gamma_geometries--------------------------------------------
%------------------------------------------------------------------------------------------
\subsubsection{Stability of the $\gamma$-geometries}
The stability of the $\gamma$-geometries is investigated the same way
as for the $\beta$-geometries. Equation~\eqref{eq:1_Formation_energy} is
used to calculate the formation energy per Pt containing surface dimer,
making close comparison with the $\beta$-geometries possible.\cite{fn:altformenerg} In the previous paragraph we found the inter dimer row interaction to be very small, compared to the intra dimer row interaction. This allows us to compare the $\gamma_1$-geometry to the
$\beta_{4\mathrm{x}}$-geometries, and the $\gamma_2$ and
$\gamma_3$-geometries to the $\beta_{5ud}$-geometry. The main difference between these geometries is the presence of a Ge-Ge dimer row in the $\beta$-geometries. Table~\ref{table:geom_energ} shows the formation energies for the $\gamma$-geometries to be in the range expected from the comparable $\beta$-geometries. As was shown in the
previous paragraph, the interaction between neighboring dimer rows is
relatively small for $\beta_{5ud}$-, \gt- and \gd-geometries, and a
larger part of the formation energy can be attributed to strain between
sequential Pt-Ge dimers. For the $\beta_{4\mathrm{x}}$- and
\ge-geometries the influence of the neighboring dimer rows is more
important. Although the geometry of the Pt-Ge dimer in the \ge-and
\bvu-geometry is nearly indistinguishable, and thus quite different
from that of the one in the \bvd-geometry, the formation energy is
about the average of that of the \bvu-and the \bvd-geometry. The
geometries of the Pt-Ge dimers in the \bf- ,\gt- and \gd-geometries are
almost identical and the angle between the antiphase Pt-Ge
dimers in a dimer row is in each case about $12\mathrm{^{\circ}}$ with
the Pt atoms on the geometrically lower side of the dimer. Furthermore,
looking at the positions of sequential dimers in a dimer row, a shift
orthogonal to the dimer row is observed. This shift increases the
distance between the Pt atoms with $0.56$--$0.61$ \AA\ in the direction
orthogonal to the dimer row. It is larger than any of such shifts
produced with normal Ge-Ge dimers in the dimer row, indicating the
crucial role of the Pt atoms.
\begin{figure}[!tbp]
\begin{center}
\includegraphics[width=8.5cm,keepaspectratio]{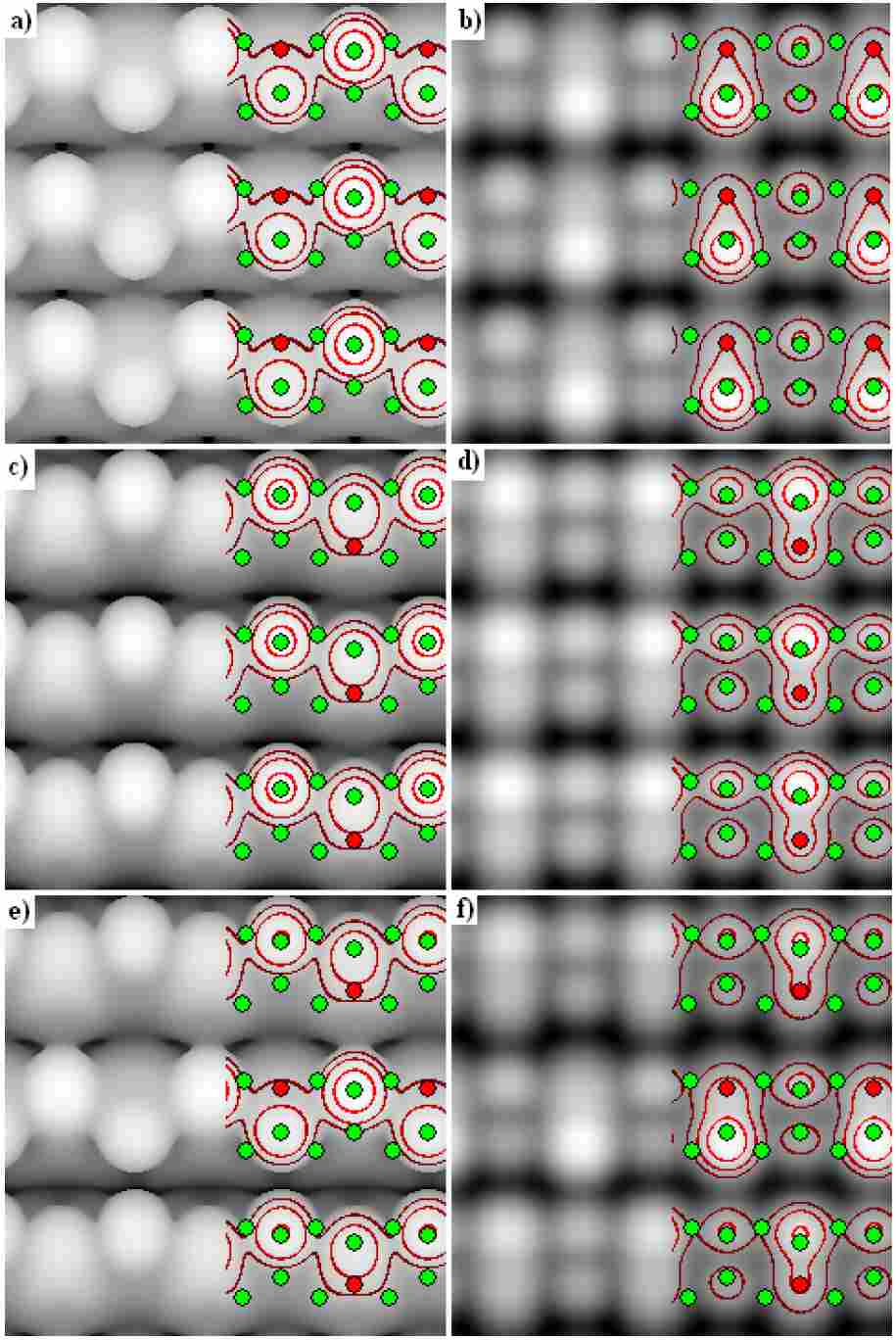}
\end{center}
\caption{(color online) Pseudo-STM images showing filled
(a, c, e) and empty (b, d, f) states of the \btu- (a, b), \btd- (c, d),
and \bz- (e, f) geometries. For the filled-state images
$\varepsilon=\varepsilon_F-0.70$ eV and for the empty-state images
$\varepsilon=\varepsilon_F+1.50$ eV is used. The constant density is chosen such that the maximum $z$ is at $3.0$ \AA\ above the highest
atom. The small dark gray (red) filled circles represent the Pt atoms in
the top layer, the small light gray (green) circles represent the positions
of the Ge atoms in the two top layers. Contours are added to indicate the
main features discussed in the text.}\label{fig:STMgeninfo}
\end{figure}
%------------------------------------------------------------------------------------------
%------------------------------Electronic_structure----------------------------------------
%------------------------------------------------------------------------------------------
\subsection{Electronic structure}\label{ssc:theorres_STM}
%% electronic structure aka pseudoSTM
%------------------------------------------------------------------------------------------
%------------------------------Electronic_structure:_beta_geometry-------------------------
%------------------------------------------------------------------------------------------
\subsubsection{General properties of the $\beta$-geometries}
%%% general knowledge
Since there is no clear evidence based on energetics alone to put
forward a single reconstruction as \textit{the} $\beta$-terrace, a more
direct comparison with the STM observations is needed. This is done by
calculating the local density of states (LDOS) $\rho(\mathbf{r},\varepsilon)$. \cite{Tersoff:prb85} The theoretical (integrated) LDOS is calculated as $\overline{\rho}(\mathbf{r},\varepsilon)\propto
\int_{\varepsilon}^{\varepsilon_F}\rho(\mathbf{r},\varepsilon')d\varepsilon'
$, with $\varepsilon_F$ the Fermi energy. A surface of constant density
is then constructed according to $\overline{\rho}(x,y,z,\varepsilon)=C$, with C a real constant. This construction gives the height $z$, as a function of the lateral position ($x,y$), which is mapped linearly onto a gray scale. This mimics the presentation of STM images in the usual constant-current mode of operation.\\
\indent Figure~\ref{fig:STMgeninfo} shows pseudo-STM images of the
\btu-, \btd-, and \bz-surface for both filled and empty states. The
filled state image of the \btu-surface shows a bright peak above
the up-atom of the Ge dimer and a slightly less bright peak above
the Ge atom of the Pt-Ge dimer. This creates a zigzag pattern of
bright spots (of two different intensities). Comparison of the
filled-state pictures of the \btu- and \btd-surface, shows images
for the Pt-Ge dimers that are clearly different. Where there is a
clear peak when the Pt atom was located between up-Ge atoms (\btu)
there is a relatively evenly spread rectangular feature when the
Pt atom is located between down-Ge atoms (\btd). This feature is
spread out over both atoms of the Pt-Ge dimer and is slightly
higher at the position of the Ge atom. The difference in height of the feature at the ($x,y$) positions of the dimer atoms is five times smaller than the difference in height between similar points of the feature produced by the Ge dimer. The Pt-Ge feature, in the \btd\ filled-state picture, has a height equal to the average height of the Ge dimer and does not protrude as far in the
trough between the dimer rows as the Ge dimer. This gives the overall impression of `battlements' (\textit{cf.}\ all dimer rows in Fig.~\ref{fig:STMgeninfo}c and the upper and lower dimer row in Fig.~\ref{fig:STMgeninfo}e).\\
\indent The \bz-surface contains both a Pt atom between up-Ge
atoms and one between down-Ge atoms. As could be expected from the
filled-state images of the \btu- and \btd-surfaces the
filled-state image of the \bz-surface shows a zigzag row and a
battlement row alternatingly. A further study of the pseudo-STM
images of the other $\beta$-geometries shows these
same structures reappearing in different combinations.\\
%With the exception of %\be\ ,$\beta_4$ and \bf\ ,
\begin{figure}[!t]
\begin{center}
    \includegraphics[width=8.50cm,keepaspectratio]{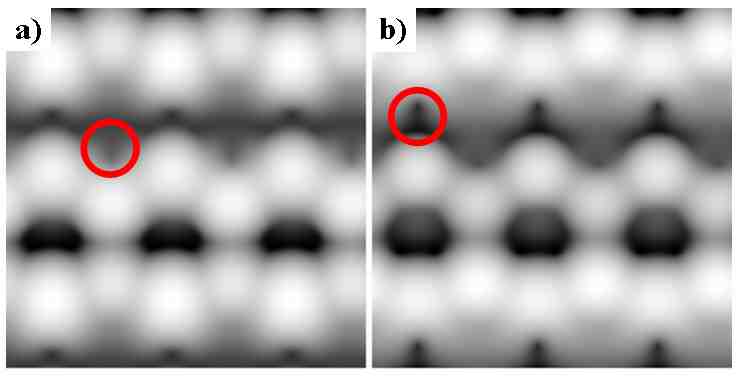}
\end{center} \caption{Pseudo-STM images showing filled (a) and empty
(b) states of the \bz-geometry close to the Fermi level, with $\varepsilon=\varepsilon_F\pm0.30$ eV. The maximum $z$ is at
$4$ \AA\ above the highest atom. There is a clear difference between
the dimer rows with the Pt atoms located between up-Ge atoms (middle
zigzag row) and the dimer rows with the Pt atoms located between
down-Ge atoms (top and bottom row). A notch (circle) in the constant density
surface above the location of the Pt atoms is visible for both the
filled- and empty-state image.}\label{fig:bet7nearF}
\end{figure}
\indent A Pt atom located between up-Ge atoms causes a feature which is
about three times as steep as the feature caused by a Pt atom between
down-Ge atoms. The Pt atom is also always located at the lower end of
these features, even though sometimes the geometry shows the Pt atom to
be the higher atom of the dimer. For the
$\beta_{4\mathrm{x}}$-geometries, the dimer row of the Pt-Ge dimers
shows a $b(2\times1)$ structure while a $(2\times2)$ structure is
found in case of the \bf-geometry. This Pt-Ge dimer row is always much
dimmer than the Ge dimer row, but it still shows nicely distinct dimer
patterns. Another careful look at the pseudo-STM structure of a Pt-Ge
dimer with the Pt atom between up-Ge atoms shows that unlike the
Ge-dimer this structure has a dim tail directed toward the Pt atom.
Although this is best visible in the empty-state pictures of Fig.~\ref{fig:STMgeninfo}, it is present for both filled- and
empty-state images.\\
\indent The above analysis of the pseudo-STM images shows that the
filled-state images of single dimers can be used as building
blocks to reconstruct the geometry of the
$\beta$-terrace (this will be shown in Sec.~\ref{sec:b6teras}).\\
\indent The empty-state pseudo-STM images for a bias far above the Fermi level
show a very simple picture, with little or no dependence of the
position and size of the Pt atom features on the neighboring Ge atoms. The Ge dimer shows up as a set of two separate peaks, while the Pt-Ge dimer
is smeared out, showing a bright ball on the side of the Ge atom and a
dim wide tail on the side of the Pt atom. Though the Pt-Ge dimer images
look very similar in the empty-state pseudo-STM pictures, they are not
exactly the same. The empty-state picture of \bz\ in Fig.~\ref{fig:STMgeninfo} shows clearly that the Pt-Ge dimer with Pt in the
up position is slightly brighter and has a somewhat more egg-like
shape. Close to the Fermi level, the pseudo-STM images of filled and
empty states are nearly indistinguishable and some interesting features
appear in the pseudo-STM images. The most eye-catching feature is the
notch in the isosurface, indicated by a circle in Fig.~\ref{fig:bet7nearF}a and b. It is caused by the Pt atom at that location. Though clearest for the Pt
atom between up-Ge atoms in a filled-state images and clearest for the
Pt atom between down-Ge atoms in the empty-state images, both are
present at either side of the Fermi level and were found in all
geometries. Checking contour plots of cross-sections in the $z$
direction at the position of these notches, showed they are caused by
the gap between two lobes of the $d$-orbitals of the Pt atom
underneath.
\begin{figure}[!t]
\begin{center}
    \includegraphics[width=8.50cm,keepaspectratio]{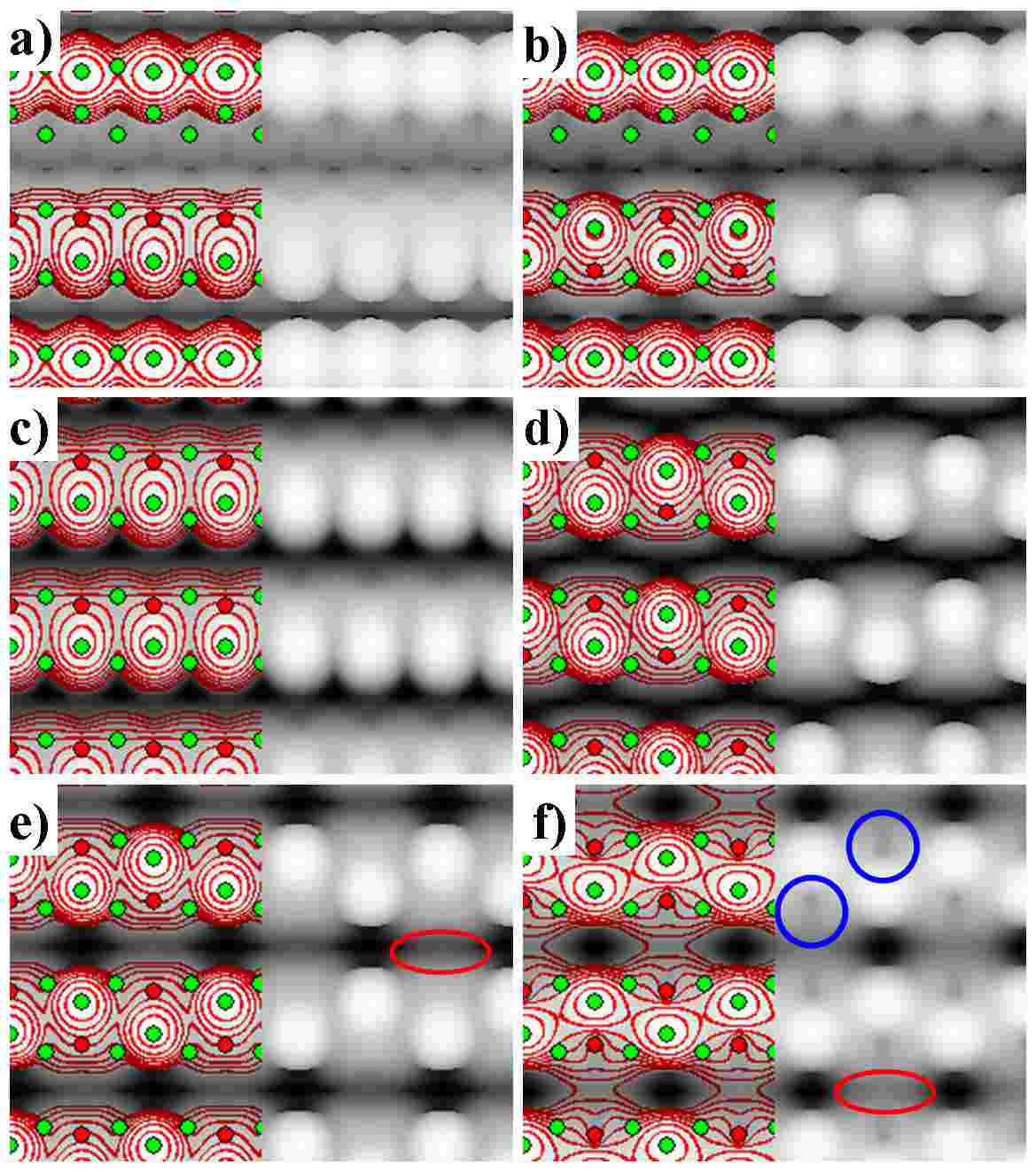}
\end{center}
\caption{(color online) Comparison of $\beta$ and $\gamma$ structures. Filled state pseudo-STM images of the $\beta_{4u}$- (a), $\beta_{5ud}$- (b), $\gamma_{1}$- (c),$\gamma_{2}$- (d), and $\gamma_{3}$-structure (e) with $\varepsilon=\varepsilon_F-0.70$ eV and an empty state pseudo-STM image of the $\gamma_{3}$-structure (f) close to the Fermi level with $\varepsilon=\varepsilon_F+0.30$ eV. The maximum $z$ is at
$3$ \AA\ above the highest atom. The green/red (light/dark gray) discs indicate the positions of the Ge (Pt) atoms in the two top layers of the system. Contours separated $0.2$ \AA\ are added to guide the eye. The circles indicate the `notch', while the ellipse shows the position of the `bridge'. The Ge dimer row in (a) and (b) show slightly broader and higher dimer images than the Pt-Ge dimer row.
}\label{fig:3b_STMgamma}
\end{figure}
%------------------------------------------------------------------------------------------
%------------------------------Electronic_structure:_gamma_geometry------------------------
%------------------------------------------------------------------------------------------
\subsubsection{General properties of the $\gamma$-geometries}
The pseudo-STM pictures of the $\gamma$-geometries show in general
the same image for the Pt-Ge dimer rows as was seen for the
corresponding Pt-Ge dimer rows in the $\beta_{4\mathrm{x}}$ and
\bf\ pictures (\textit{cf.}\ Fig.~\ref{fig:3b_STMgamma}). For the \ge-geometry, the pseudo-STM pictures show
dimer rows with a bright spot at the Ge atom position. This
spot becomes sharper when the bias becomes smaller, both above and
below the Fermi energy. The \gt- and \gd-geometries show nearly
identical pictures (\textit{cf.}\ Fig.~\ref{fig:3b_STMgamma}d and e), the main difference being the phaseshift between neighboring dimer rows, which causes the appearance of a very dim (at least $1.5$ \AA\ lower than the dimer peaks in the pseudo-STM pictures) bridge between the Pt atoms of neighboring
dimer rows, indicated with an ellipse in Fig.~\ref{fig:3b_STMgamma}e and f. This bridge is most clear in the empty state pictures.
Again it is the Ge side of the Pt-Ge dimers which lights up very
bright, causing a zigzag pattern of bright spots along each dimer
row. For small biases this zigzag pattern becomes more pronounced
and also the notch above the Pt atoms appears, as indicated with the circles in Fig.~\ref{fig:3b_STMgamma}f. At a large negative bias the bright spot has a dim tail directed toward the Pt atom.  All Pt-Ge dimer images for \gt\ and \gd\ are identical, unlike for the $\beta$-geometries.
\begin{figure}[!t]
\begin{center}
   \includegraphics[width=8.250cm,keepaspectratio]{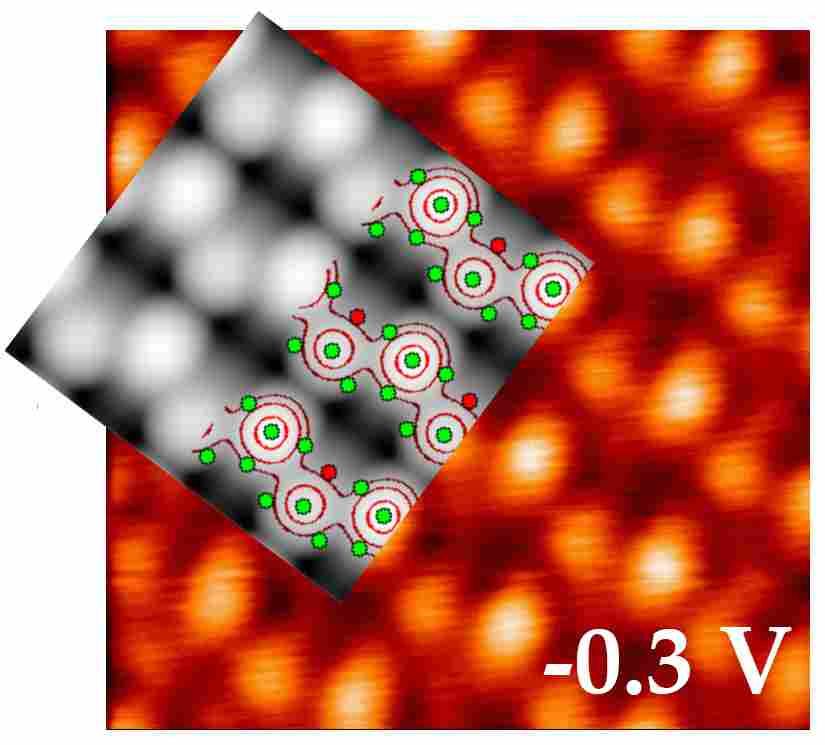}
\end{center}
\caption{(color online) Comparison of the filled-state pseudo STM image of the
\bsugeom (gray scale overlay) with the corresponding experimental STM
image of the $\beta$-terrace. The STM image is obtained using a sample
bias of $-0.3$ V, while the pseudo STM image is generated using
$\varepsilon=\varepsilon_F-0.70$ eV and a maximum $z$ of $4$ \AA\ above
the highest atom. The small red (dark gray) discs represent the positions
of the Pt atoms in the top layer, while the green (light gray) discs
represent the positions of the Ge atoms in the top two layers. The QDRs
are clearly visible in both STM images, as is the presence of two
different types of dimer.}\label{fig:beta6filled}
\end{figure}
%------------------------------------------------------------------------------------------
%------------------------------Electronic_structure:_beta6u_=_beta-terrace-----------------
%------------------------------------------------------------------------------------------
\subsubsection{\bsu-structure as geometry for the $\beta$-terrace}\label{sec:b6teras}
\begin{figure}[!t]
\begin{center}
  \includegraphics[width=8.250cm,keepaspectratio]{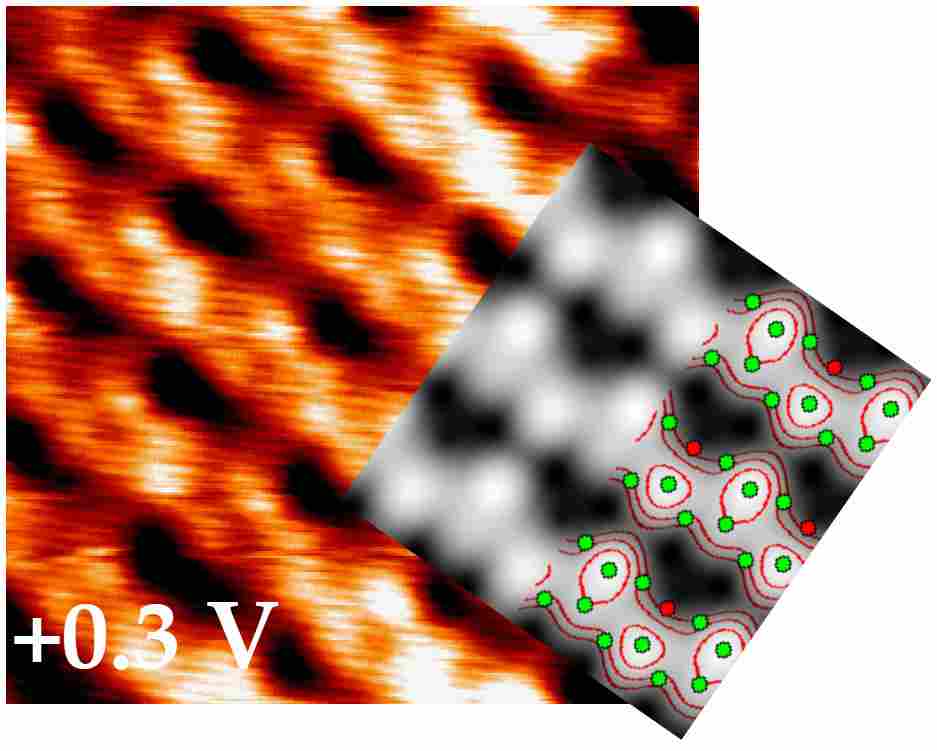}
\end{center}
\caption{(color online) Same as Fig.~\ref{fig:beta6filled} but now for the empty-state
images. The sample bias used for the STM image is $+0.3$ V and
$\varepsilon=\varepsilon_F+0.70$ eV is used for the pseudo STM
image. Note the triangular holes above the Pt atoms (red discs) which are present in both the theoretical and the experimental STM image. This is a unique feature for the $\beta_{6u}$-structure.\label{fig:beta6empty}}
\end{figure}

\begin{figure}[!t]
\begin{center}
    \includegraphics[width=8.50cm,keepaspectratio]{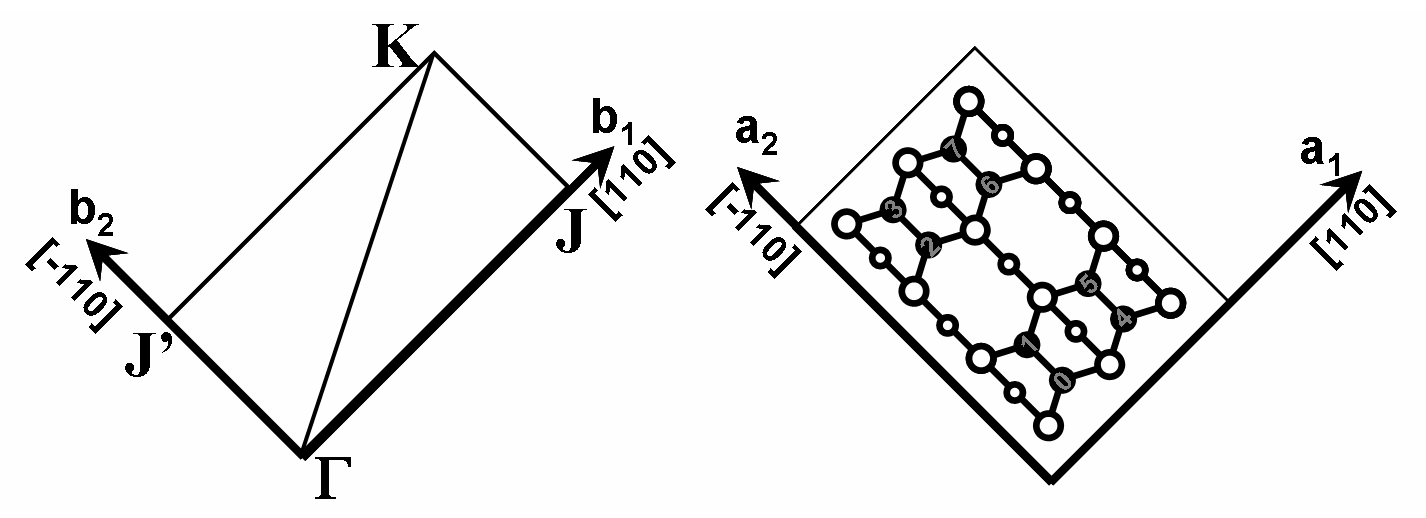}\\
\end{center}
  \caption{High symmetry lines and points of the surface Brillouin zone (left) for the $\beta$-structures. A surface unit cell is shown on the right, indicating the orientation of the dimer rows.}\label{fig:6_0SBZ_b6u}
\end{figure}

\begin{figure}[!t]
\begin{center}
    \includegraphics[width=8.0cm,keepaspectratio]{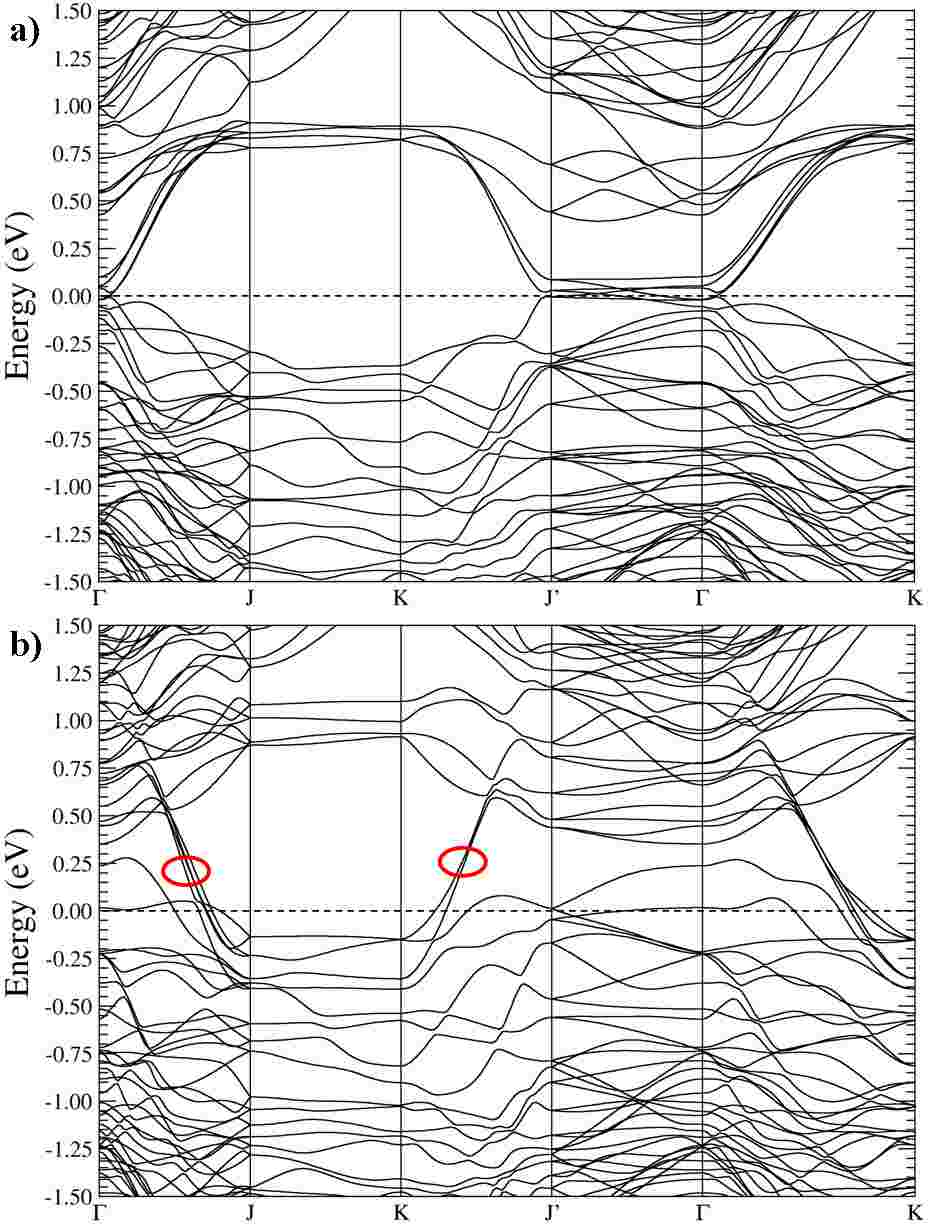}\\
\end{center}
  \caption{Comparison of the band structure, along high symmetry
  lines (\textit{cf.}\ Fig.~\ref{fig:6_0SBZ_b6u}), of the Ge(001) $c(4\times2)$ reconstruction (a) and the \bsugeom
  (b). The energy zero is set at the Fermi level.}\label{fig:10bands_b6u}
\end{figure}

%%% full circle identification of beta6
In this paragraph we will show that, although most $\beta$-structures
studied are energetically available, only the \bsugeom can be related
to the experimentally observed $\beta$-terrace. Figures~\ref{fig:beta6filled} and \ref{fig:beta6empty} show (experimental) STM
images of the $\beta$-terrace at biases of $-0.30$ V and $+0.30$ V. The
filled-state image clearly shows two types of dimers alternating in
each dimer row. One of the dimer types shows up as a big bright spot on
one side of the dimer row while the other dimer type shows up as a
smaller somewhat dimmer spot on the other side of the dimer row, with a
dim tail extending to the opposite side of the dimer row. These are the
only two types of dimers appearing in the image. Comparing this to the
results of the pseudo-STM images seen in the previous paragraph, the
first type of dimer is the Ge-Ge dimer while the second type of dimer
presents all the typical features of the Pt-Ge dimer with the Pt atom
located between up-Ge atoms. This allows us to exclude all
$\beta$-geometries containing Pt at a down location, leaving only the
\btu-, \bvu-, and \bsu-geometry. Taking the relative ordering of the features of neighboring dimer rows into account, the geometry which represents the
$\beta$-terrace can only be the \textit{\bsu-geometry}. Filled- and
empty-state pseudo-STM images of the \bsugeom are displayed on top of
the experimental pictures in Fig.~\ref{fig:beta6filled} and
\ref{fig:beta6empty}.
The empty-state pseudo-STM image (Fig.~\ref{fig:beta6empty}) shows some
very specific features of this geometry. There are \textit{triangular
holes} where the Pt atom is located and the Ge atoms of the top layer
also create a triangular feature dominated by the zigzag structure of
the up-Ge atoms and the Ge atoms in the Pt-Ge dimers. Comparison of the
empty-state images in Fig.~\ref{fig:beta6empty} reveals the existence
of these features in both images, further strengthening our belief that
the \bsu-geometry is the geometry of the $\beta$-terrace. Another look
at Table~\ref{table:geom_energ} also confirms this geometry to be more
stable than the Ge $b(2\times1)$ surface reconstruction. Furthermore,
for the $\beta_{6\mathrm{x}}$-geometries, the Pt atoms are also
distributed in the most homogeneous way.
The pseudo-STM images show only for the \bsu-geometry this level of agreement with the experimental STM images, allowing the \bsu-structure to be presented as \textit{the} geometry of the experimentally observed $\beta$-terrace.\\
\indent Figure~\ref{fig:10bands_b6u}b shows the band structure of the
$\beta_{6\mathrm{u}}$-geometry along the high symmetry lines
$\Gamma$-J-K-J$^{\prime}$-$\Gamma$-K of the surface BZ (\textit{cf.}\ Fig.~\ref{fig:6_0SBZ_b6u}).
Figure~\ref{fig:10bands_b6u}a shows the band structure of the Ge(001)
c($4\times2$) reconstruction along the same lines for comparison. In
general, the two band structures show similar behavior. Perpendicular
to the dimer row direction, along the J-K and J$^{\prime}$-$\Gamma$
line, the bands show little dispersion. Along the J$^{\prime}$-$\Gamma$
line, a few of these bands cross the Fermi level. The main difference
between the two band structures is located along the $\Gamma$-J and
K-J$^{\prime}$ lines. For the \bsu-structure we find bands with a
strong dispersion opposite to what is seen for the Ge(001) surface. These bands are indicated with the ellipses in Fig.~\ref{fig:10bands_b6u}b.
Furthermore, these bands also cross the Fermi level in the second half
of the $\Gamma$-J line and in the first half of the K-J$^{\prime}$ line, resulting in  the metallic behavior observed
for the $\beta$-terrace. \cite{Gurlu:thesis} Taking a closer look at
these specific bands near the crossing of the Fermi level, we find that
their character contains three main components. The largest component
has a Pt-like character. The other two (smaller) components have a
Ge-like character: more specifically, the second layer Ge atoms bound
to the Pt atoms in the top layer and the Ge up-atom of the Ge surface dimer that is located in between the Pt atoms. The orbital character of the Pt contribution is mainly
$p_x,\ p_y$ and $d_{xz},\ d_{yz}$, making it highly planar. The Ge
contribution depends on the Ge atoms involved. The orbital character of
the contribution of second layer Ge atoms is mainly $sp^3$, while the
top layer Ge atoms bound to the Pt atom contribute mainly a $p_z$
orbital character, indicating the presence of a $\pi$-bond with energy
close to the Fermi level for the Pt-Ge dimer. Interestingly, the Ge
up-atoms of the Ge dimers in between Pt atoms along the dimer row have a significant contribution to most of the bands we are interested in. The orbital character in this case is mostly $p_x$ and $p_y$. The presence of bands
close to (and crossing) the Fermi level connected to an orbital
character directed along the dimer row direction, indicates the
possible presence of surface conduction channels along this direction.
These bands could be the source of the confined states observed in
Ref.~\onlinecite{Oncel:prl05} and \onlinecite{Houselt:nanol06}.\\
\begin{figure}[!t]
\begin{center}
  \includegraphics[width=8.0cm,keepaspectratio]{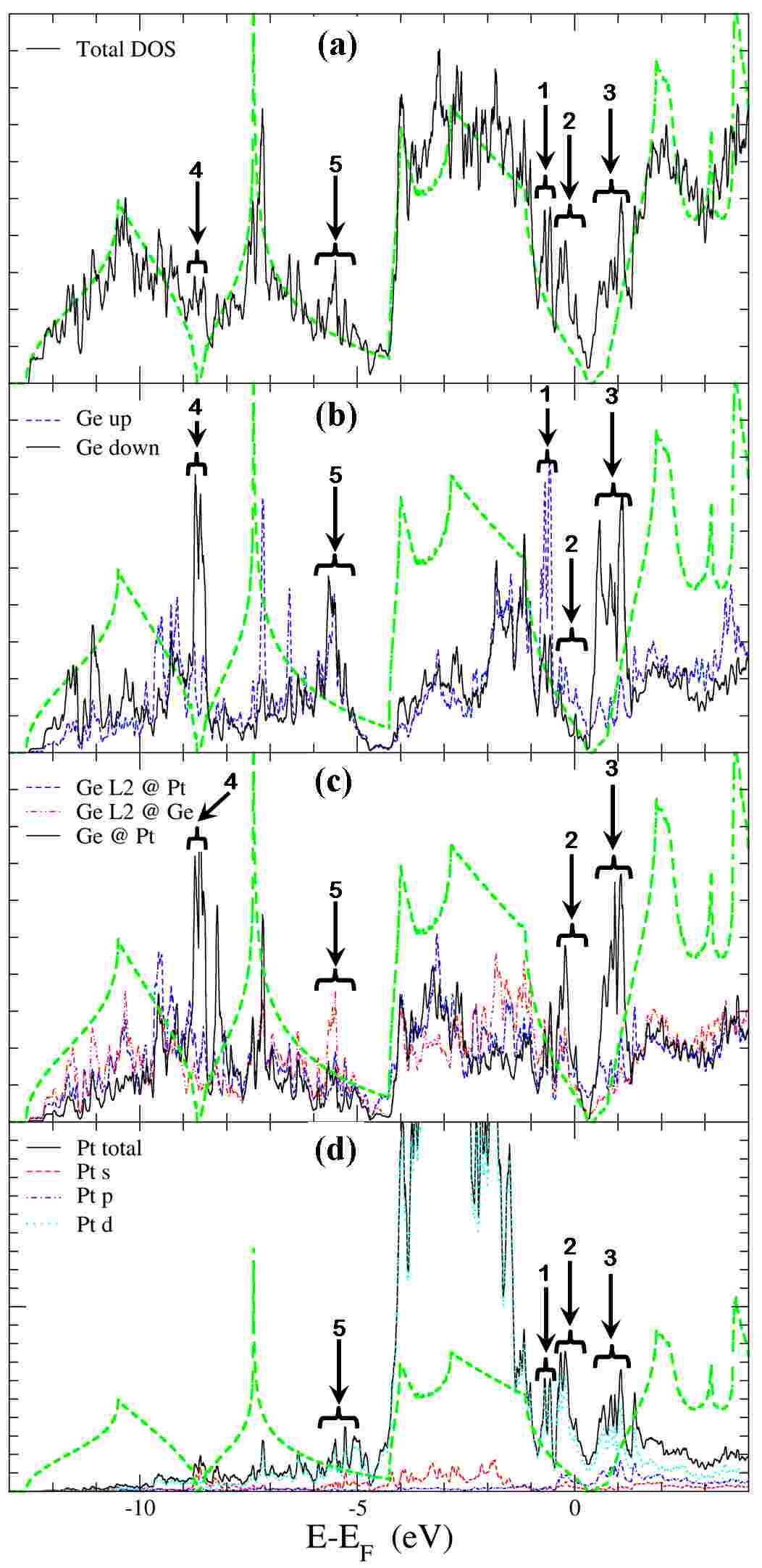}\\
\end{center}
  \caption{(color online) Total DOS of the \bsugeom and the LDOS of the
  surface atoms. The bold dashed green curve shows the DOS of Bulk Ge,
  it is shifted to align the edges of the BG regions. The labeled peaks are discussed in the text.}\label{fig:11DOS_b6u}
\end{figure}
\indent Figure~\ref{fig:11DOS_b6u} shows the density of states (DOS) of the \bsugeom and the LDOS of the surface atoms.\cite{fn:Ge_BG}
The total DOS roughly follows the DOS observed for bulk Ge, as could be
expected. The main difference is found in the band gap (BG) region. Here, three
distinct peaks can be observed, indicated by the labels $1$, $2$, and $3$ in Fig.~\ref{fig:11DOS_b6u}. The two outer peaks ($1$ and $3$) are also observed
for a reconstructed Ge(001) surface, where they are linked to the
dangling bonds of the Ge surface dimer atoms. Since half the surface
dimers of the \bsugeom are Ge dimers, it is not surprising to observe
these states. Figure \ref{fig:11DOS_b6u}b
shows the LDOS of the up and down atoms of the Ge surface dimers.
Comparison to the Ge(001) c$(4\times2)$ and b($2\times1$)
reconstruction, shows the peak due to the Ge-down atom ($3$) to be quite
broad, with a width comparable to what is found for the b($2\times1$)
reconstruction. This broadening probably originates from the small
tilt-angle of the Pt-Ge dimers. This makes the angle between subsequent dimers in a quasi-dimer row (QDR) smaller, making it more comparable to the zero angle of the b($2\times1$) reconstruction. The Ge-up atom, shows two peaks in the BG ($1$ and $2$ in Fig.~\ref{fig:11DOS_b6u}b). The strong peak, labeled $1$ is also found for the Ge-up atom of the Ge(001) c$(4\times2)$ and b($2\times1$) reconstructions. The weaker second peak has a non-zero value for the DOS at the Fermi level and is one of the contributions to peak $2$ in the total DOS.\\
\indent The LDOS image of the Ge atoms bound directly to the Pt atom
shows something interesting (\textit{cf.}\ Fig.~\ref{fig:11DOS_b6u}c). The Ge atom of the Pt-Ge dimer also has two peaks in the BG region, contributing to peaks $2$ and $3$ in the total DOS. A sharp peak around the same value as seen
for the Ge down atom at the conduction band (CB) side ($3$), and a second
peak near the valence band (VB) side of the BG region ($2$). This second peak is located just at the Fermi level. Between these two peaks the LDOS goes to zero, just above the Fermi level.\\
\indent The second layer Ge atoms (indicated with L$2$ in Fig.~\ref{fig:11DOS_b6u}) show roughly the same structure, both have a small contribution to the conducting peak $2$ of the total DOS. However, in the
BG region it are only the second layer Ge atoms bound to the Pt atoms of the
top layer that have non-zero values over the entire BG region (\textit{cf.}\ blue dashed curve in Fig.~\ref{fig:11DOS_b6u}c). Second layer Ge atoms bound to the Ge atoms of the top layer have a LDOS which goes to zero. The LDOS for the Pt atoms, presented in Fig.~\ref{fig:11DOS_b6u}d, shows the presence of Pt $d$ states at and around the Fermi level, with $3$ peaks at the same locations as found for the Ge atoms bound to these Pt atoms. For the Ge atoms, it are mainly $p$ states contributing to these peaks near the Fermi level, showing clearly the presence of a hybridization of Ge $p$ with Pt $d$
states. Together with the Ge up-atoms and Pt atoms, the Ge atoms bound to the
Pt atoms form a corridor of `metallic' atoms along the edge of the QDR, giving rise to the metallic peak $2$ of the total DOS, in agreement with the experimental observations.\cite{Gurlu:thesis}\\
\indent The additional peaks, $4$ and $5$, in the total DOS are related to the surface Ge atoms, and are also observed for the Ge(001) c$(4\times2)$ and b($2\times1$) reconstructions.\\
%------------------------------------------------------------------------------------------
%-------------Electronic_structure:_the_other_beta_and_gamma_in_experiments?---------------
%------------------------------------------------------------------------------------------
\subsubsection{Experimental existence of other $\beta$-geometries}
\indent Although the comparison between experimental and pseudo-STM images in
Fig.~\ref{fig:beta6filled} and \ref{fig:beta6empty} shows clearly that
the clean $\beta$-terrace has a \bsu-geometry, the formation energies
given in Table~\ref{table:geom_energ} indicate that also other
geometries should be observed experimentally. In particular the $\beta_{4u}$ structure has a slightly lower energy. The latter structure leads to STM images that are completely different to what is observed experimentally, as do other non $\beta_{6u}$ structures. Our calculations therefore indicate that the formation of such structures is kinetically hindered. Investigation of
experimental STM images has shown the existence of (small) regions
where dimer rows contain a mix of different (non \bsu) geometries.
These geometries were mostly found near defects in the normal
$\beta$-terrace, such as dimer vacancies and the endpoints of a
NW. Because these geometries seem related to defects, we will not
go into the details of their geometry and focus solely on the \bsu-geometry.\\
%%\indent However, the question why only the $\beta_{6u}$-geometry is observed %%experimentally in the $\beta$-terraces remains.
%------------------------------------------------------------------------------------------
%-----------------------------Electronic_structure:_beta6u_geometry:whole_9_yards----------
%------------------------------------------------------------------------------------------
%% more info on beta6
\subsection{The \bsu-geometry}\label{ssc:theorres_beta6u}
\begin{table*}[!t]
\center{\textbf{Bond lengths and dimer tilt angle.}}
\begin{ruledtabular}
\begin{tabular*}{\textwidth}{l|ccccccccccccccc}
&$\omega$&A&B$_1$&B$_2$& C$_1$& C$_2$& C$_3$& C$_4$ &d$_1$&d$_2$& d$_3$& t$_1$& t$_2$& t$_3$ & t$_4$ \\
\hline
Ge b$(2 \times 1)$ & 19.42 & 2.450 & 2.512 & 2.426 & 2.475 & 2.449 & 2.543 & 2.475 & 0.815 & 0.775 & 0.099 & 3.818 & 0 & 0 & 0 \\
Ge c$(4 \times 2)$ & 19.60 & 2.511 & 2.506 & 2.430 & 2.463 & 2.481 & 2.481 & 2.462 & 0.842 & 0.749 & 0.007 & 3.817 &0.283 & 0.335 & 0.332 \\
\bsu\ Ge-dimer     & 20.82 & 2.462 & 2.552 & 2.433 & 2.479 & 2.435 & 2.492 & 2.469 & 0.875 & 0.731 & 0.027 & 3.764 &0.153 & 0.049 & 0.003 \\
\bsu\ Pt-dimer     & $-4.25$ & 2.354 & 2.378 & 2.521 & 2.479 & 2.435 & 2.492 & 2.469 & 0.174 & 1.358 & 0.027 & 3.764 &0.153 & 0.049 & 0.003 \\
\end{tabular*}
\end{ruledtabular}
\caption{Bond lengths and distances in \AA\ and the dimer tilt angle
$\omega$ in degrees. All as defined in Fig.~\ref{fig:bondangledef}.
}\label{table:bonds}
\end{table*}
\begin{figure}[!t]
\begin{center}
  \includegraphics[width=8.50cm,keepaspectratio]{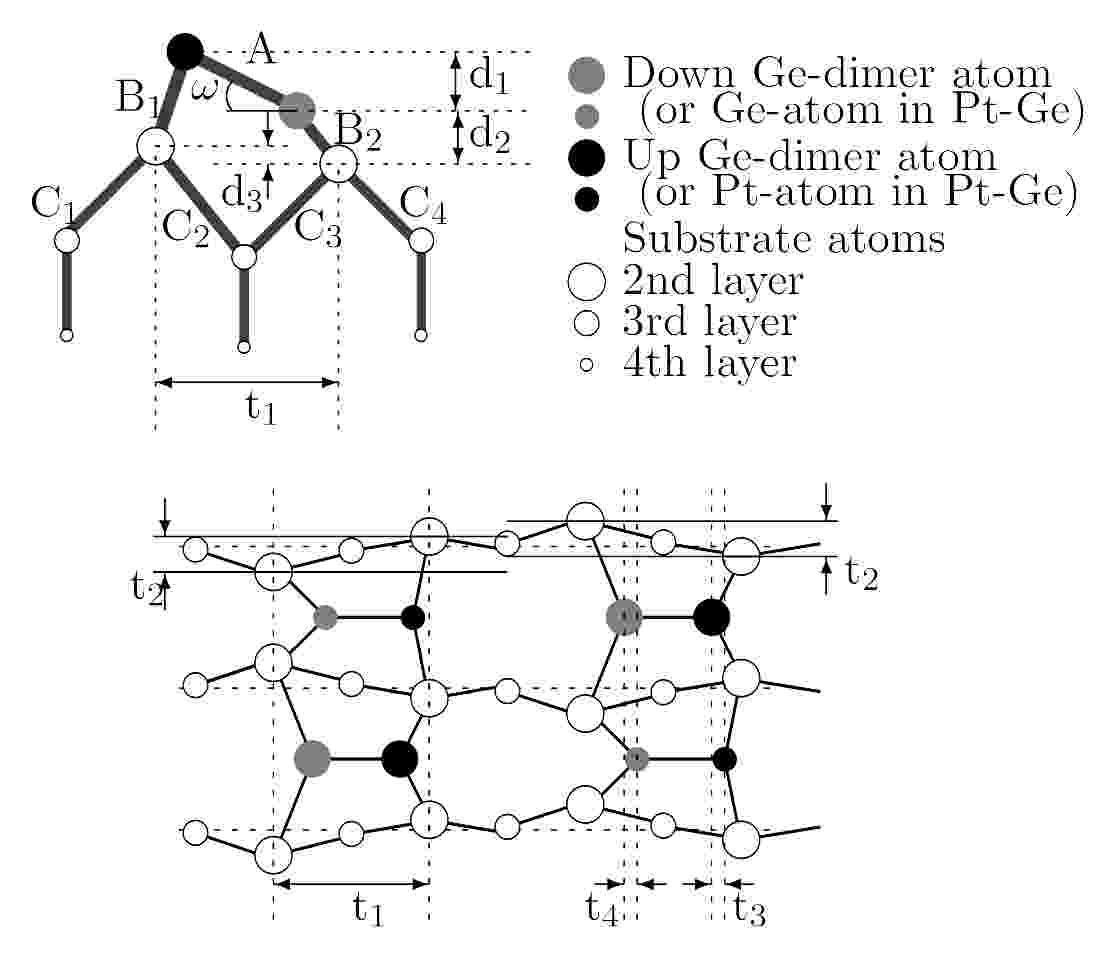}\\
\end{center}
  \caption{Top: Diagram showing a side view of a relaxed Ge-Ge
dimer, including the labeling used for both Ge-Ge and Ge-Pt surface
dimers.\newline Bottom: Top view schematic representation of the
$\beta_{6u}$-reconstruction.}\label{fig:bondangledef}
\end{figure}
%over de beta6u
\indent To characterize the geometry of the $\beta_{6u}$ structure, it is compared to the geometry of the Ge(001) $b(2\times1)$ and $c(4\times2)$ reconstructions. This shows a slight shift of the second and third layer Ge atoms of the $\beta_{6u}$ structure in the surface plane with regard to the Ge $b(2\times1)$ equilibrium positions. In what follows we will show that this deformation of the geometry can be understood as a combination of three geometric modifications, each linked to a symmetry breaking of the top layer structure.\\
\indent In the second layer of the $\beta_{6u}$ geometry, two inequivalent sites are present: i) a Ge atom between a top layer Ge up atom and a Pt atom, and ii) a Ge atom between a top layer Ge down atom and a Ge atom of a Pt-Ge dimer. The Ge atom at the latter site is displaced toward the neighboring Ge atom of the Pt-Ge dimer, while the Ge atom at the former site is displaced toward the neighboring Ge up atom (\textit{cf.}\ Fig.~\ref{fig:bondangledef}). In comparison to the Ge $b(2\times1)$ reconstruction, in the surface plane these atoms appear as twisted around the in-plane position of the third layer Ge atom in between them. For this reason, such a modification of the geometry, \textit{i.e.} the twisting of the tetrahedral structure, is known as \textit{twisting}.\cite{Gay:prb99}\\
\indent The Ge $(2\times2)$ surface reconstruction, for which a diagrammatic representation can be found in figure $2$ of Ref.~\onlinecite{Gay:prb99}, is a simple Ge surface reconstruction which only shows the \textit{twisting} modification with regard to the $b(2\times1)$ reconstruction. \textit{Twisting} is believed to be a strain release mechanism, where strain (orthogonal to the surface) due to different surface dimer buckling is (partially) compensated by an in-plane modification of the second layer geometry, indicated with t$_2$ in Fig.~\ref{fig:bondangledef}. This means that the \textit{twisting} parameter t$_2$ is related to the relative buckling angle of sequential surface dimers in a dimer row. Assuming a linear relation as first order approximation one finds $\mathrm{t}_2\sim\mid\omega_{\mathrm{dimer1}}-\omega_{\mathrm{dimer2}}\mid=\Delta\omega$, where $\omega_{\mathrm{dimerX}}$ is the buckling angle $\omega$ shown in Table~\ref{table:bonds} for each of the dimers involved. Because the relative
angle between sequential dimers can be used as a measure for the \textit{twisting}, it becomes possible to estimate the \textit{twisting} for similar surfaces via the ratio of their relative buckling angle
\begin{equation}\label{eq:1_simpleRatio}
R=\frac{\mathrm{t^{rec1}_{2}}}{\mathrm{t^{rec2}_{2}}}=\frac{\Delta\omega^{\mathrm{rec1}}}{\Delta\omega^{\mathrm{rec2}}},
\end{equation}
where rec$1$ and rec$2$ refer to each of the two reconstructions being compared.\\
\indent Unlike the \textit{twisting} observed for Ge surface reconstructions with differently buckled surface dimers, the \textit{twisting} found in the $\beta_{6u}$ geometry is asymmetric. From Fig.~\ref{fig:bondangledef} follows that t$_2$ can be considered to consist of two parts; the component along the dimer row direction of the bond C$_2$ and the component along the dimer row direction of the bond C$_3$. The ratio of the contributions of C$_2$ and C$_3$ to t$_2$ is \mbox{$1:1$} for the Ge $(2\times2)$ and Ge $c(4\times2)$ reconstructions, while it is $6.3:9$ for the $\beta_{6u}$ reconstruction. This means that, in comparison to the Ge $c(4\times2)$ reconstruction, the second layer Ge atom between the Pt atom and the Ge up atom is pulled slightly back toward the Pt atom in the surface plane.\\
\indent Using Eq.~\eqref{eq:1_simpleRatio} to calculate the ratio $R$ of the \textit{twisting} of the $\beta_{6u}$ reconstruction to the Ge $c(4\times2)$ reconstruction making use of the values for t$_2$ presented in Table~\ref{table:bonds} we find a value of $0.5406$ for $R$. However, making use of the relative buckling angle we find a value of $0.6395$ for $R$. Unlike the Ge $c(4\times2)$ reconstruction, there are different kinds of surface dimers on the $\beta_{6u}$ surface. To include this inhomogeneity of the surface dimers in Eq.~\eqref{eq:1_simpleRatio}, we add weight factors to the angle contributions. We found the mass percentages of the dimers to be good weight factors. This leads to a \textit{twisting} ratio for the $\beta_{6u}$ to the Ge $c(4\times2)$ structure:
\begin{eqnarray}%\label{eq:2_t2_angleratio}
R&=&\frac{\Big|\Big(\frac{\mathrm{M}_{\mathrm{Ge}}+ \mathrm{M}_{\mathrm{Pt}}}{3\mathrm{M}_{\mathrm{Ge}} +\mathrm{M}_{\mathrm{Pt}}}\Big)\omega_{\beta_{6u} \mathrm{Pt}} -\Big(\frac{2\mathrm{M}_{\mathrm{Ge}}}{3\mathrm{M}_{\mathrm{Ge}} +\mathrm{M}_{\mathrm{Pt}}}\Big)\omega_{\beta_{6u}\mathrm{Ge}}\Big|}{\omega_{c(4\times2)}} \label{eq:2_t2_angleratio} \\
&=&\frac{\mathrm{t}_2^{\beta_{6u}}}{\mathrm{t}_2^{\mathrm{Ge}
c(4\times2)}}, \nonumber
\end{eqnarray}
with M$_{\mathrm{x}}$ the atomic mass of a single (average) atom of type x, and $\omega_{\mathrm{y}}$ the angle $\omega$ shown in Table~\ref{table:bonds} for the specific dimer y. Using Eq.~\eqref{eq:2_t2_angleratio} a value of $0.5142$ is found for the ratio R, in agreement with the value obtained for the ratio of the t$_2$ parameters. Also for other $\beta$ structures, equations equivalent to  Eq.~\eqref{eq:2_t2_angleratio} give an improvement of the estimated ratios of t$_2$ over the ratios obtained through Eq.~\eqref{eq:1_simpleRatio}. \textit{E.g.}\ for the $\beta_{1ud}$ structure compared to the Ge $c(4\times2)$ structure, the ratio of the \textit{twisting} parameters t$_2$ is $0.5477$. Using the relative buckling angles in Eq.~\eqref{eq:1_simpleRatio} one finds a value of $0.0354$, which is much to small. However, using the weighted buckling angles, as in Eq.~\eqref{eq:2_t2_angleratio}, one finds a value of $0.5121$, in good agreement with the ratio of the parameters t$_2$.\\
\indent This shows t$_2$ to be a useful parameter for comparison of the asymmetry induced strain in these systems. As a result, the difference of the \textit{twisting} parameter, t$_2$, between the Ge \cge\ and \bsurec can be attributed to the introduced Pt atoms. On the one hand by leveling out the Pt-Ge dimer, \textit{i.e.} the $\omega$ term, and on the other hand through the weight factor.\\
\indent In conclusion, \textit{twisting} is the result of the relative buckling angle of sequential dimers in a dimer row which is influenced strongly by the Pt atoms present in the $\beta_{6u}$ structure.\\
\begin{figure}[!t]
\begin{center}
  \includegraphics[width=8.50cm,keepaspectratio]{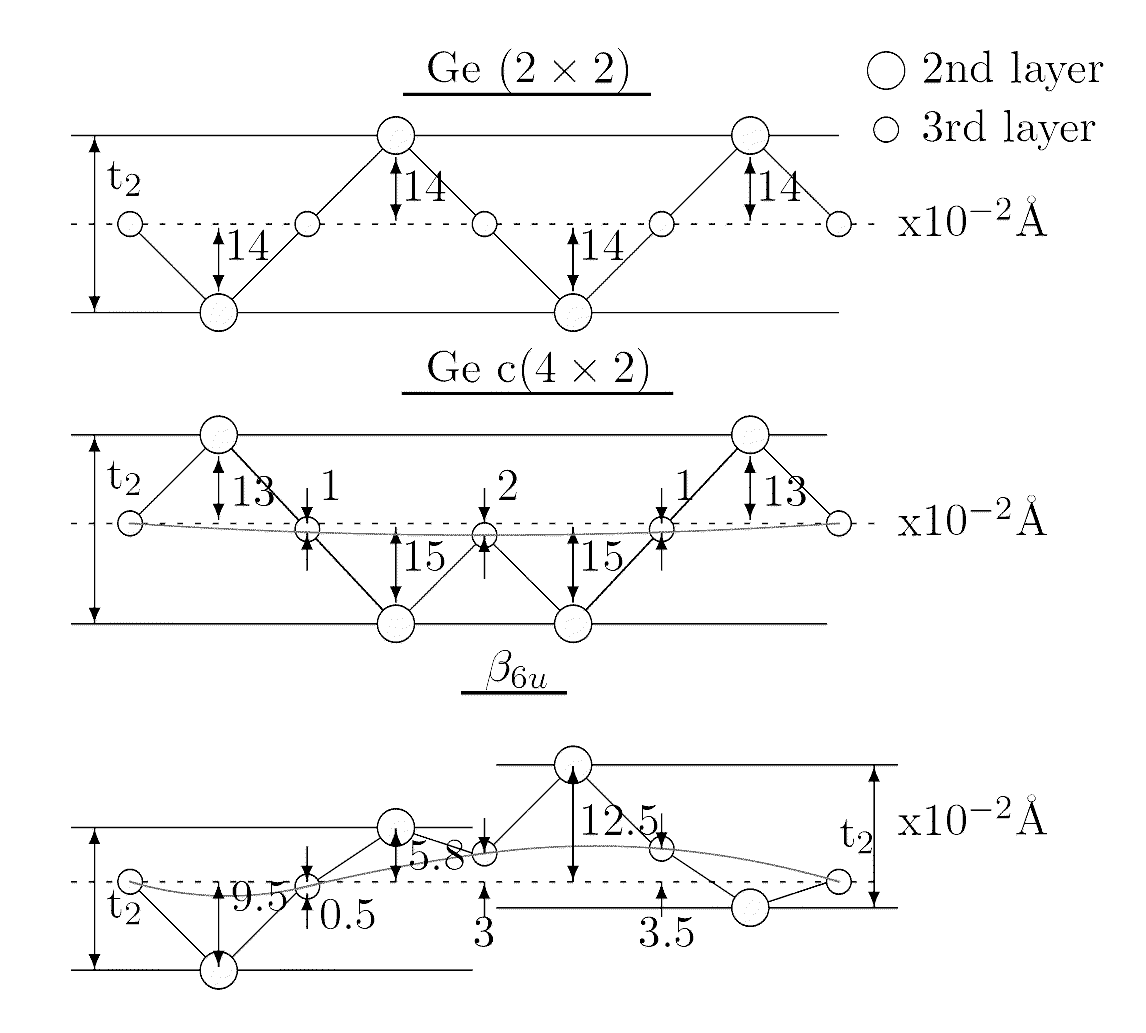}\\
\end{center}
\caption{The diagrams show a top view of the upper row, as represented in the
diagram of the \bsurec in Fig.~\ref{fig:bondangledef}, of second and
third layer Ge atoms in a Ge$(2\times2)$-, Ge $c(4\times2)$- and
$\beta_{6u}$-reconstruction. The adjacent top layer atoms, as represented in the bottom
diagram of Fig.~\ref{fig:bondangledef}, are then from left to
right:
Ge $(2\times2)$: Ge up-dimer atom, down-dimer atom,
up-dimer atom and down-dimer atom.
Ge $c(4\times2)$: Ge
down-dimer atom, up-dimer atom, up-dimer atom and down-dimer
atom.
\bsu: Ge atom, Pt atom, Ge down-dimer atom and Ge
up-dimer atom.
(Data for the Ge $(2\times2)$-reconstruction is
taken from Gay \textit{et al.}\cite{Gay:prb99})}\label{fig:twistcomp}
\end{figure}
\indent Figure~\ref{fig:twistcomp} gives diagrams of the top view of a row of second and third layer Ge atoms in geometries with
increasing complexity. The horizontal dashed lines give an arbitrary
zero-line with regard to which all amplitudes are given. The Ge
($2\times2$)-reconstruction shows the simple image of \textit{twisting}
of the tetrahedral arrangement, discussed above. The image of the Ge $c(4\times2)$ reconstruction presents, in addition to \textit{twisting}, a \textit{bending} of the entire second and third layer structure, indicated with the gray curve in Fig.~\ref{fig:twistcomp}. It shows a slight displacement of of the third layer Ge atoms in the plane along the dimer row direction. This is due to the asymmetry between the C$_1$, C$_4$ and the C$_2$, C$_3$ bonds as a result of the opposite buckling of dimers in adjacent dimer rows.\\
\indent In the $\beta_{6u}$ structure, just like the Ge $c(4\times2)$ reconstruction, the surface dimers in adjacent dimer rows show an opposite buckling which results in a \textit{bending}, as shown with the gray curve in the bottom picture of Fig.~\ref{fig:twistcomp}. Furthermore, the presence of Pt atoms in the top layer dimers seems to increase the
amplitude of the bending by a factor of two and the extrema of the
bending curve move from the third layer atoms between the dimer rows
to the third layer atoms underneath the dimer rows.\\
\indent The presence of two different dimer types in the $\beta_{6u}$ structure complicates this image even further, as can be seen in the bottom picture of Fig.~\ref{fig:twistcomp}. On top of the previous two distortions a \textit{shift} of the \textit{twisting} t$_2$ between adjacent dimer rows is present. \textit{Shift} is the displacement of the horizontal lines defining the \textit{twisting} parameter t$_2$ between neighboring dimer rows (\textit{cf.}\ bottom pictures of Fig.~\ref{fig:bondangledef} and Fig.~\ref{fig:twistcomp}), in other words; it is the \textit{shift} along the dimer row direction of the \textit{twisting} parameter t$_2$ its bounding box between adjacent dimer rows. This \textit{shift} is a consequence of the induced asymmetry in the components along the dimer row direction of the C$_2$ and C$_3$ bonds contributing to t$_2$, which in turn is due to the asymmetry of having different dimer types in a dimer row, as well as in adjacent dimer rows.\\
\indent In case of the $\beta_{6u}$ structure, the \textit{shift} is directed toward the Pt-Ge dimer, away from the Ge-Ge dimer (\textit{cf.}\ Fig.~\ref{fig:bondangledef}, bottom picture). This makes \textit{shift}, which is due to different dimer types in one system, the third surface modification present in the $\beta_{6u}$ geometry. It shrinks the surface cell, formed by the first and second layer atoms, containing the Pt-Ge dimer in favor of the Ge-Ge dimer surface cell. Table~\ref{table:bonds} also shows a decrease of the parameter t$_1$, reducing the size of the Pt-Ge dimer surface cell even further and partially compensating the expansion of the Ge-Ge dimer surface cell.\\
\indent The bond length of the Ge-dimer in the \bsu\ reconstruction is only
slightly longer than it was in the \brec\ but still significantly
shorter than it is in the $c(4\times2)$-reconstruction. The
accompanying change in buckling angle accounts for this slight
elongation. Also, the back bonds of the Ge atoms of all dimers are
longer than their counterparts in the asymmetric Ge surface
reconstructions. This while the height of these back bonds for the
Ge-Ge dimers is similar to the one found for the
$c(4\times2)$-reconstruction. t$_3$ and t$_4$, shown in  Fig.~\ref{fig:bondangledef}, give the in-plane displacement along the surface dimer bond direction between sequential dimers in a single dimer row, i.e. the alignment of the surface atoms along the dimer rows. Although they are not zero, they are at least one order of magnitude smaller than the values
obtained for the $c(4\times2)$-reconstruction, and unlike the \crec
there is an asymmetry between the two displacements. his results from the difference in length of Pt-Ge and Ge-Ge bonds.\\
\indent The C$_n$-bonds between the second and third layer Ge atoms have lost all symmetry, this is due to the two types of surface dimers present in the $\beta_{6u}$ structure. This is unlike the reconstructions studied by Gay \textit{et al.} \cite{Gay:prb99} where only one type of surface dimer is present in each system, and as a consequence the C$_n$ bonds show much symmetry.
%------------------------------------------------------------------------------------------
%-------------------------------Other_geometries-------------------------------------------
%------------------------------------------------------------------------------------------
\subsection{Other geometries}\label{ssc:theorres_othergeom}
\begin{figure}[!tbp]
\begin{center}
   \includegraphics[width=8.50cm,keepaspectratio]{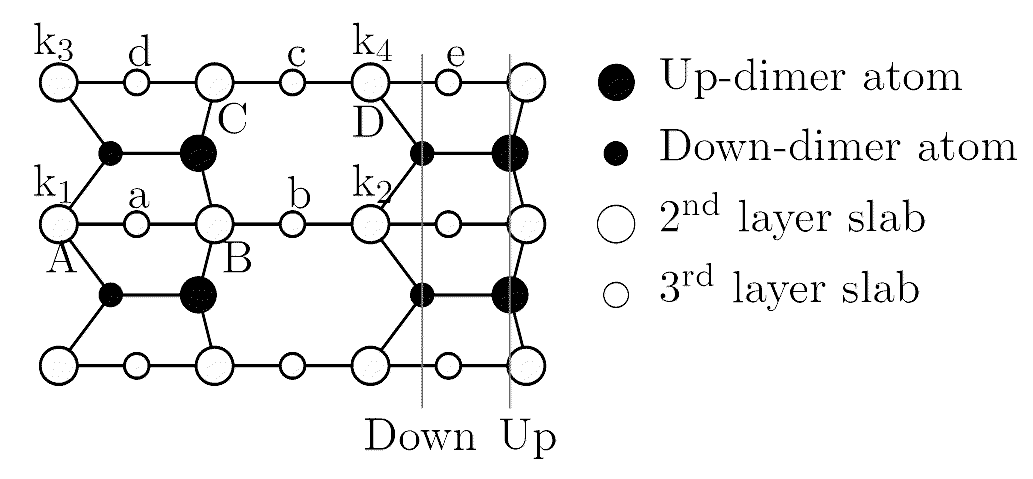}\\
\end{center} \caption{Locations of Ge atoms substituted by Pt atoms in
the second (capital) and third (small) layer. The second layer Ge atoms
pushed to the surface by third layer Pt atoms are marked by the k$_{\mathrm{x}}$.}\label{fig:l2l3nomen}
\end{figure}

%% other geometries: pt in L2 and L3
Though the previous section shows clearly that the $\beta$-terrace has
a \bsu-geometry, it is still interesting to have a look at some
geometries containing Pt atoms in the second or third layer of the
system. Previous calculations and experimental observations have
shown that Pt atoms can also be found in subsurface
positions. \cite{Gurlu:prb04}  Before the transformation to the
$\beta$-terrace, there is what G\"url\"u \textit{et al.} call an
$\alpha$-terrace. This type of terrace is characterized by a
large number of dimer vacancy defects. Also, in this initial stage of Pt growth on Ge(001) there are
still a number of features which are not entirely understood,
such as `indented dimers'.\\
\indent Figure~\ref{fig:l2l3nomen} shows the positions where Ge atoms
were substituted with Pt atoms in the second and third layer of the
system. The final geometries were obtained the same way as the
$\beta$-geometries, \textit{i.e.}\ substitution of Ge atoms by Pt atoms
in a Ge(001) b($2\times1$)-reconstructed slab and relaxation of the
geometry using the conjugate gradient method keeping the positions of
the two center layers fixed.
%------------------------------------------------------------------------------------------
%-------------------------------Other_geometries:_Pt_in_L2---------------------------------
%------------------------------------------------------------------------------------------
\subsubsection{Pt in the second layer}
When Ge atoms in the second layer are substituted with Pt atoms, there
are two possible inequivalent sites to consider. One under the down Ge
atoms and one under the up Ge atoms of the surface dimers. Table~\ref{table:formenergyl2l3} shows the formation energy per Pt atom for
these two locations differs only $64$ meV and Pt favors the position
under the up Ge dimer atoms. Table~\ref{table:formenergyl2l3} also
shows that the introduction of a second Pt atom in the second layer
improves the stability of the system considerably, increasing the
formation energy by $28$ and $123$ meV per Pt atom. This difference
from what would be expected from the single Pt substitution results,
shows there is a positive correlation between the Pt atoms in
the second layer. Furthermore, comparison of Table~\ref{table:geom_energ} and Table~\ref{table:formenergyl2l3} shows the
formation energies per surface unit are roughly in the same range as
the formation energies per surface unit of the studied $\beta$- and
$\gamma$-geometries, making these structures energetically also viable
candidates for the $\beta$-terraces. Pseudo-STM images on the
other hand clearly show this is not the case.
\begin{table}[!t]
\center{\textbf{Formation energy.}}
\begin{ruledtabular}
%sterrekes moeten $\ast$ worden
\begin{tabular}{ccrc}
\makebox[3cm]{Layer} & \makebox[3cm]{Pt position(s)} & \multicolumn{2}{c}{\makebox[3cm]{$E_f$}} \\
& & \multicolumn{2}{c}{(meV)} \\
\hline
 $2$ & A (down) & \makebox[1.5cm][r]{24}  & \\
 $2$ & B (up)   & $-$40 & \\
 $2$ & A \& D (down) & $-$4 & \\
 $2$ & B \& C (up) & $-$163 & \\
\hline
 $3$ & a (row) & $-$1115 & \\
 $3^{\ast}$(k$_1$) & a (row) & $-$155 & \\
 $3$ & b (trough)& 68 & \\
 $3$ & a \& b & $-$630 & \\
 $3^{\ast}$(k$_1$\& k$_2$) & a \& b & 467 & \\
 $3$ & a \& c & $-$467 & \\
 $3^{\ast}$(k$_1$) & a \& c & $-$66 & \\
 $3$ & a \& d & $-$616 & \\
 $3^{\ast}$(k$_1$\& k$_3$) & a \& d & $-$32 & \\
 $3$ & a \& e & $-$1141 & \\
 $3^{\ast}$(k$_1$\& k$_4$) & a \& e & $-$346 & \\
\end{tabular}
\end{ruledtabular}
\caption{Formation energy per Pt atom of reconstructions containing Pt
atoms in the second and third layer. The Pt positions are shown in Fig.~\ref{fig:l2l3nomen}. The asterisk indicates that the Ge atoms at the
$k_{\mathrm{x}}$ positions (between brackets) are removed from the
system, and the values were obtained after relaxation of the modified
system.\label{table:formenergyl2l3}}
\end{table}
\begin{figure}[!t]
\begin{center}
    \includegraphics[width=8.50cm,keepaspectratio]{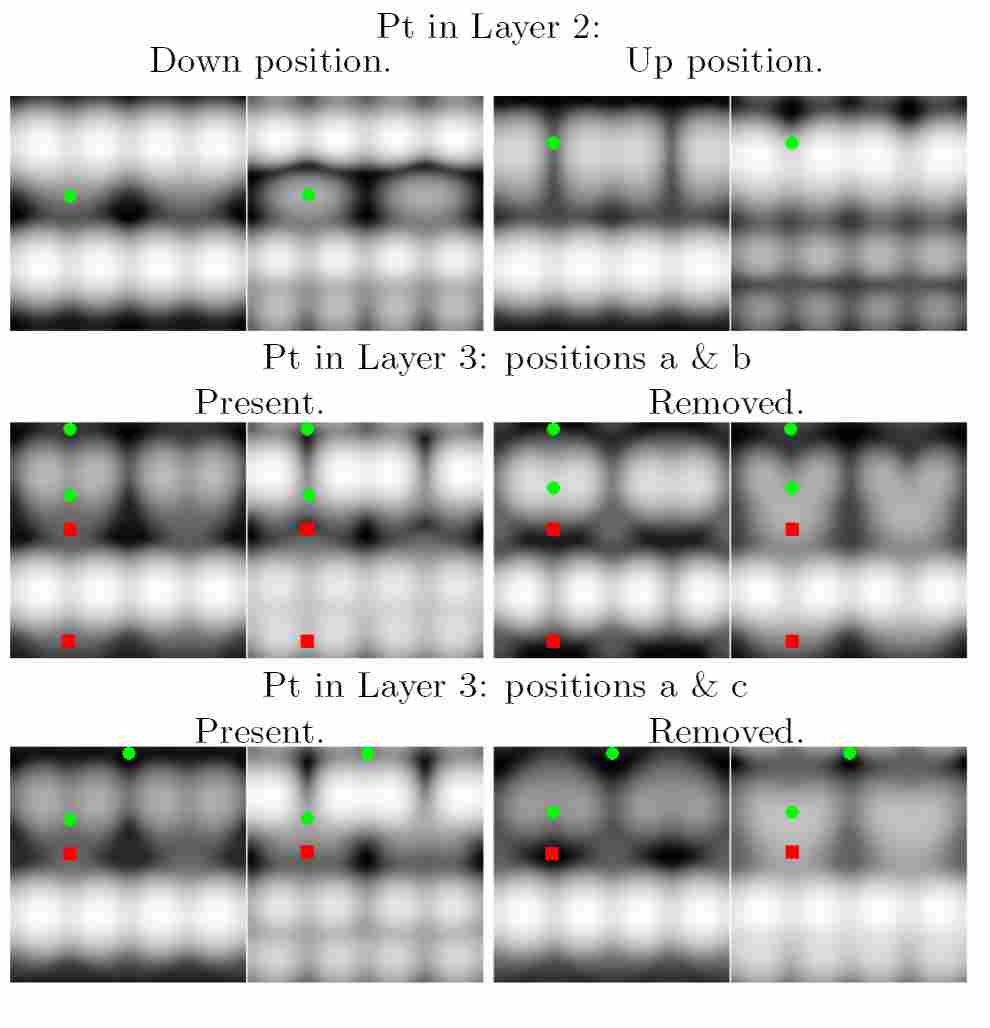}
\end{center}
\caption{(color online )Pseudo-STM images showing filled- (left of each pair) and empty- (right of each pair) state images for the geometries containing
Pt in the second and third layer. $\varepsilon=\varepsilon_F\pm0.70$ eV
is used in these simulations and the maximum $z$ is chosen at $4$
\AA\ above the highest atom. The positions of the Pt atoms are
indicated by the (green) discs and the (red) squares indicate the
positions of second layer Ge atoms which are pushed up to the first
layer level. `\emph{Present}' and `\emph{removed}' refers to these Ge
atoms being present or having been removed ($\ast$-configurations of
Table~\ref{table:formenergyl2l3}). The bottom dimer row in each picture
shows Ge dimers which are not (or barely) influenced by the Pt atoms
present. The dimers at the top of each picture show a clear
modification caused by the Pt atoms present underneath.
\label{fig:ptinl2l3}}
\end{figure}
The top part of Fig.~\ref{fig:ptinl2l3} shows the influence of the
substitution of a single Ge atom by a single Pt atom in the second
layer of the system. In both cases the images of the two Ge dimers on
top of the Pt atom merge into one image, with the
location of the overlap on the side where the Pt atom is located.\\
\indent For a Pt atom located at the down side this fusing is only
seen for small bias voltages, for large bias the images are almost
indistinguishable from the normal Ge dimer images. Also, the
brightness of the dimers is roughly the same as for the Ge dimers
on a normal Ge(001) surface.\\
\indent In case the Pt atom is located at the up side of the dimer row,
the position where the two images of the Ge dimers merge moves from
being on top of the Pt atom, for large negative biases, to being on top
of the Ge atom in between two Pt atoms, for small negative biases and
all positive biases. Also, in this case there is a clear difference in
brightness between the normal Ge dimers and the dimers on top of a Pt
atom. For negative biases, the normal Ge dimers are most bright while
for positive biases, the modified dimers are brighter. The brightest
part of the modified dimers extends over the entire dimer for small
biases, while for large positive (and negative) bias the brightest part
of the image is at the side of the up Ge atoms. This gives the
impression of a dimer located at one side of the dimer row, and
directed along this dimer row. This is only an impression because these
structures are formed by the bending, toward and away from the Pt atom,
of the Ge $p$ orbitals. When all Ge atoms at the up side in the second
layer under a dimer row are substituted by Pt atoms, the images of the
above Ge dimers do not merge anymore. Their image only lengthens and
becomes a long oval shape which is dimmer than the normal Ge dimer at
negative bias and brighter than the normal Ge dimer at positive bias.
This elongated shape persists at positive bias such that there is a
clear difference between this image and the image of a normal Ge dimer,
for which the single dimer image is replaced by two spots, one for each
Ge atom.
%------------------------------------------------------------------------------------------
%-------------------------------Other_geometries:_Pt_in_L3---------------------------------
%------------------------------------------------------------------------------------------
\subsubsection{Pt in the third layer}
Substitution in the third layer can also be done at two inequivalent
sites. The first is under the dimer row and the second is in the
trough, between dimer rows. Table~\ref{table:formenergyl2l3} shows a
clear preference for Pt to be positioned under a dimer row, while a Pt
atom in the trough position is less favorable than the original
b($2\times1$)-reconstruction. This again shows the preference of Pt to
be surrounded by Ge atoms. This is understandable since the
Pt-Ge bond is energetically favorable over a Pt-Pt or a Ge-Ge bond.
Table~\ref{table:formenergyl2l3} also shows that the introduction of a
second Pt atom in the third layer about halves the formation energy per
Pt atom, if the second Pt is located at a next nearest neighbor position.
Though this might be expected for a combination of one row and one
trough positioned Pt atom, it is also the case for two Pt atoms in row
positions. Unlike Pt in the second layer, Pt in the third layer prefers
to be positioned far from other Pt atoms. Positions $a$ and $e$ seem to
be separated enough to remove this negative correlation. Comparing
Table~\ref{table:geom_energ} and Table~\ref{table:formenergyl2l3} shows
clearly that Pt under a dimer row is energetically the most favorable
configuration, with a formation energy about one order of magnitude
larger than the other configurations, in agreement with previous
calculations. \cite{Gurlu:apl03}\\
\indent Indeed by comparing Tables~\ref{table:geom_energ} and~\ref{table:formenergyl2l3} it can be observed that thermodynamically Pt atoms prefer to substitute Ge atoms in the third layer. However, this substitution involves breaking the local, bulk-like Ge structure, whereas substitution of Ge by Pt in the top surface layer involves breaking Ge-Ge surface bonds. From experiments on the Ge($001$) surface it is known that at the anneal temperature of $\sim1050$ K, the surface Ge dimers break up.\cite{Zeng:ss02} At this temperature the bulk is still intact, as melting starts at a much higher temperature. This should favor the incorporation of Pt in the top surface layer. Indeed, experiments show that the formation of the beta-terrace is connected to the anneal temperature of $\sim1050$ K.\cite{Gurlu:apl03,Gurlu:thesis}\\
\indent Geometrically the most impressive change is caused by a Pt atom under a dimer row at site a (or an equivalent site). The atom pushes the second layer Ge atom (in case of site a, indicated with $k_{\mathrm{1}}$ in Fig.~\ref{fig:l2l3nomen}) at the down side of the
row up to the first layer deforming the surface geometry. Also in the case where the geometry contains two Pt atoms, at positions $a$ and $b$, there is an additional lifting of Ge atom $k_2$, which might be considered caused by the Pt atom at position $a$. The Pt atom (site a) pushes Ge atom $k_1$ up to a first layer position, but it also pushes the Ge atom (site B) in between the two Pt atoms a little bit away. This Ge atom in turn pushes against the second Pt atom (site b) which then lifts Ge atom $k_2$ creating a chain effect in which a
horizontal displacement is transformed into a vertical one. Table~\ref{table:formenergyl2l3} shows the removal of these Ge atoms (after
relaxation to a new equilibrium geometry) costs about $0.6$ to $1.1$ eV
per Ge atom removed. So if the Ge atoms are to be considered as ejected
atoms which move away to a position at a step edge, then the energy
gained from placing a Pt atom in a row position is almost entirely lost
and we again end up in the energy range of the $\beta$- and
$\gamma$-geometries. The exception being the geometry placing Pt atoms
in positions $a$ and $e$, which resembles the $\beta_6$-geometry, and
thus could be considered a possible starting position for the
$\beta_{6u}$-geometry.\newline
\\
\indent The influence on the pseudo-STM pictures is similar as seen for
the Pt atoms in the second layer. The charge lobes of the up Ge dimer
atoms are pulled toward one another over the Pt atom at negative bias
while for positive bias they are pushed away. For the down Ge dimer
atoms, the charge lobes are always pulled toward one another over the
Pt atom. The merging of the dimer images in the pseudo-STM pictures of
Fig.~\ref{fig:ptinl2l3} show this nicely. A more important effect
caused by this deformation of the charges is the dimming at negative
and the brightening at positive biases of the nearby Ge dimers, which
can be substantial, as shown in Fig.~\ref{fig:ptinl2l3}. This indicates
a charge transfer from Ge to Pt atoms. Removal of the $k_{\mathrm{x}}$
Ge atoms forces the down Ge atoms to make a new bond, which generally
is with a third layer atom, either the Pt atom under the dimer row or
the Ge atom in the trough. This removal causes a lowering of the dimer
structure into the surface consequently causing a lowering of the
electronic structure and hence dimming the dimer image in the
pseudo-STM pictures even further. The change in the $z$-position of the
dimer atoms is not the only geometrical effect caused by this removal.
The dimers also bend toward one another over the location of the
removed Ge atom. Consequently the two dimer images in the pseudo-STM
picture merge into one single heart-shaped feature, of which the
orientation depends on the bias, as can be seen in Fig.~\ref{fig:ptinl2l3}.
%------------------------------------------------------------------------------------------
%-------------------Other_geometries:_crouching_atom,_hidden_dimer-------------------------
%------------------------------------------------------------------------------------------
\subsubsection{Missing dimers or hidden dimers?}
In the previous two paragraphs it was shown that the introduction
of Pt in the second or third layer causes significant changes in
the electronic structure of the surface. The difference in
height/brightness of the dimer images in the pseudo-STM pictures strongly
depends on the simulated bias. Combining the visual information
from the STM-image with the height differences of the maxima at
different biases (both positive and negative) should make it possible to clearly distinguish the different
geometries containing Pt in the second and third layer in experiment.\\
\indent We generated pseudo-STM images for all the structures containing Pt in the second or third layer, using simulated biases of $-1.50$, $-0.70$, $+0.70$, and $+1.50$ V to study the influence of varying bias. For each bias, a set of STM images was generated using different values for $z$ ($2.5$ \AA, $3.0$ \AA, and $4.0$ \AA) to construct surfaces of different constant current, allowing us to study the effect of varying current. In real STM experiments, there are other variables that influence the final STM image, such as the tip geometry. As a consequence, one should not interpret the values given below as absolutes, but instead as relative values indicating the trends which should be observed in experiment for these structures.\\
\indent For a Pt atom in the second layer at the down location,
the maxima calculated for dimers on top of the Pt atom and dimers
without Pt nearby, are roughly the same (less than $0.1$ \AA).
Here the image at positive bias would give the telltale signature
with the dimmed oval structure over a pair of down Ge dimer atoms,
which is $\sim$0.45 \AA\ lower at $+0.70$ V and $\sim$0.3 \AA\
lower than the maxima at $+1.50$ V bias.\\
\indent For second layer Pt atoms at the up location there is the
merging of two dimer images into one single rectangular feature,
which at low negative bias will shift one full dimer position with
regard to the positive bias position. Furthermore, at increasing
positive bias the maximum will seem to move to one side of the
dimer row. At negative bias, the maximum of the structure will be
$\sim$0.25 \AA\ lower than that of a normal Ge dimer, with this
height difference becoming larger at smaller bias. At positive
bias on the other hand the maximum will be about $0.3$~\AA\ higher than that of a normal Ge dimer.\\
\indent Pt in the third layer also modifies the electronic
structure in a recognizable way. Pt under the dimer row lowers the
peaks of the dimer row above by $0.5$ to $0.7$ \AA\ at negative
bias, while a Pt atom in the trough modifies the Ge dimer, with
its up Ge atom nearby, lowering the maxima of these dimers by only
$\sim$0.25 \AA. For the latter, these dimers are also slightly
more asymmetric than the normal ones and at small positive bias
the dimer image is still one feature while for a normal dimer the
image splits up into two separate atom images. This splitting
still occurs for the modified dimer but at a larger positive bias.
At positive bias the modified dimers are highest for both Pt
positions. For low positive biases they are $\sim$0.25 \AA\
higher, than the normal dimers, while for large positive bias this
difference shrinks to only $\sim$0.1 \AA. Removal of the Ge atoms,
which were pushed to the top layer, changes the electronic
structure dramatically from the normal Ge dimer image. In each
case two dimer images are merged into one single large
heart-shaped feature which is about $0.6$ to $0.7$ \AA\ lower than
the maxima of the normal dimers at negative bias, and about $0.8$ \AA\ lower at small positive bias. At large positive bias this
difference reduces to $\sim$0.25 \AA\ at which point the maximum
of the structure is located at one side of the dimer row, giving
the feature a somewhat butterfly-like appearance.\\
\indent Combinations as seen in Fig.~\ref{fig:ptinl2l3} give similar
results. For example, the height difference of the dimer images, for Pt
in the third layer at positions $a$ and $c$ in the filled-state image
is $\sim$0.8 \AA\ and in the empty-state image this is $\sim$0.2 \AA.
When the $k_1$ Ge atom is removed, the filled-state image now gives a
height difference of $\sim$1.0 \AA\ while the empty-state image shows
one of $\sim$0.6 \AA. For both
filled- and empty- state images the normal dimers are highest.\\
\indent Although we used a very simple model to calculate the
pseudo-STM pictures, the values obtained suggest that some of
these geometries could cause experimentally observed features in
the $\alpha$-terrace, such as indented dimers or two dimer
vacancies. Indented dimers are a structure two dimers long and based on the
above results we conclude that this feature is caused by a Pt atom
in a second layer up position or a third layer trough position.
Two dimer vacancies on the other hand might not be actual missing
dimers but merely due to the charge transfer from Ge-Ge dimers to
a Pt atom in a third layer row position underneath. At this moment
the above remains uncertain. A high resolution STM study over
varying bias of these defects however, should make a conclusive
statement on this subject possible.
%------------------------------------------------------------------------------------------
%-------------------Other_geometries:_crouching_atom,_hidden_dimer-------------------------
%------------------------------------------------------------------------------------------
\section{Conclusions}\label{sc:conclusions}
%Conclusion
We study the possible geometries containing $0.25$
and $0.5$ ML of Pt substituted in the top layer of a Ge(001)
reconstructed surface. Zero temperature \textit{ab initio} calculations
show that the formation of Pt-Ge dimers constitutes the most favorable
reconstructions. They also show the spontaneous formation of
alternate buckling in most geometries due to the introduction of Pt.
Comparison of calculated STM images with experimental observations is
necessary to show the $\beta$-terrace has a \bsu\ geometry, which has
some similarities with the Ge(001) c($4\times2$) reconstruction. It is shown that the Pt-Ge dimers are tilted in anti-phase to the Ge-Ge dimers, and a
thorough description the \bsu\ geometry is given in terms of \textit{twisting}, due to the anti-phase buckling of dimers in a dimer row, \textit{bending}, due to the anti-phase buckling of dimers in adjacent dimer rows, and \textit{shift}, due to the asymmetry of having two types of dimers present on the surface.
Additionally, we study geometries containing Pt atoms substituted in
the second and third layer and show the most stable configuration to
be one with Pt atoms located under the Ge dimer row. It is also shown
that Pt atoms in this position push a second layer Ge atom to the
surface and that the removal of this atom does not increase the
formation energy but instead decreases it. In all these `other'
geometries, with exception of those where Ge atoms were removed, the
presence of Pt atoms both modifies and dims the Ge dimers on top for
negative bias and brightens them for positive bias. The height
differences we find indicate that these other geometries could be
responsible for experimentally observed indented dimers and two-dimer
vacancies. Also, the calculated STM images of these other geometries
show clearly that they can not be responsible for the $\beta$-terrace,
though some are energetically more favorable, indicating that the
$\beta_{6u}$-geometry is a local minimum.

\section*{Acknowledgements}
%Special thanks etc
%----------------------------------------------------------------------
We acknowledge Prof.\ Harold Zandvliet and Dr.\ Arie van Houselt for many useful discussions, and for making available their experimental data.
This work is part of the research program of the ``Stichting voor
Fundamenteel Onderzoek der Materie" (FOM) and the use of
supercomputer facilities was sponsored by the ``Stichting
Nationale Computer Faciliteiten" (NCF), both financially supported
by the ``Nederlandse Organisatie voor Wetenschappelijk Onderzoek"
(NWO).
%----------------------------------------------------------------------

%*************************************************************************************************************
% BIBLIOGRAPHY
%*************************************************************************************************************
% Put in \nocite{*} so all entries in the bibliography are included
%\nocite{*}
% This GATHER command is useful for when you want to
%use WinEdt's Gather functionality, i.e., type
% \cite{} and a popup box appears with all of your
%citations to choose from.  Leave the % on the next line.

% The commented way of writing Gather is the only correct way of doing it !!!
%GATHER{cmsdanny.bib}
%GATHER{betaterasnotes.bib}
%\bibliography{cmsdanny,betaterasnotes}

\begin{thebibliography}{32}
\expandafter\ifx\csname natexlab\endcsname\relax\def\natexlab#1{#1}\fi
\expandafter\ifx\csname bibnamefont\endcsname\relax
  \def\bibnamefont#1{#1}\fi
\expandafter\ifx\csname bibfnamefont\endcsname\relax
  \def\bibfnamefont#1{#1}\fi
\expandafter\ifx\csname citenamefont\endcsname\relax
  \def\citenamefont#1{#1}\fi
\expandafter\ifx\csname url\endcsname\relax
  \def\url#1{\texttt{#1}}\fi
\expandafter\ifx\csname urlprefix\endcsname\relax\def\urlprefix{URL }\fi
\providecommand{\bibinfo}[2]{#2}
\providecommand{\eprint}[2][]{\url{#2}}

\bibitem[{\citenamefont{Barth et~al.}(2005)\citenamefont{Barth, Constantini,
  and Kern}}]{Barth:nat05}
\bibinfo{author}{\bibfnamefont{J.~V.} \bibnamefont{Barth}},
  \bibinfo{author}{\bibfnamefont{G.}~\bibnamefont{Constantini}},
  \bibnamefont{and} \bibinfo{author}{\bibfnamefont{K.}~\bibnamefont{Kern}},
  \bibinfo{journal}{Nature} \textbf{\bibinfo{volume}{437}},
  \bibinfo{pages}{671} (\bibinfo{year}{2005}), \bibinfo{note}{and references
  therein.}

\bibitem[{\citenamefont{Lin et~al.}(1993)\citenamefont{Lin, Wan, Glueckstein,
  and Nogami}}]{LinXF:prb93}
\bibinfo{author}{\bibfnamefont{X.~F.} \bibnamefont{Lin}},
  \bibinfo{author}{\bibfnamefont{K.~J.} \bibnamefont{Wan}},
  \bibinfo{author}{\bibfnamefont{J.~C.} \bibnamefont{Glueckstein}},
  \bibnamefont{and} \bibinfo{author}{\bibfnamefont{J.}~\bibnamefont{Nogami}},
  \bibinfo{journal}{Phys. Rev. B} \textbf{\bibinfo{volume}{47}},
  \bibinfo{pages}{3671} (\bibinfo{year}{1993}).

\bibitem[{\citenamefont{Kageshima et~al.}(2001)\citenamefont{Kageshima, Torii,
  Tano, Takeuchi, and Kawazu}}]{Kageshima:ss01}
\bibinfo{author}{\bibfnamefont{M.}~\bibnamefont{Kageshima}},
  \bibinfo{author}{\bibfnamefont{Y.}~\bibnamefont{Torii}},
  \bibinfo{author}{\bibfnamefont{Y.}~\bibnamefont{Tano}},
  \bibinfo{author}{\bibfnamefont{O.}~\bibnamefont{Takeuchi}}, \bibnamefont{and}
  \bibinfo{author}{\bibfnamefont{A.}~\bibnamefont{Kawazu}},
  \bibinfo{journal}{Surf. Sci.} \textbf{\bibinfo{volume}{472}},
  \bibinfo{pages}{51} (\bibinfo{year}{2001}).

\bibitem[{\citenamefont{Shimakura et~al.}(1998)\citenamefont{Shimakura, Minoda,
  Tanishiro, and Yagi}}]{Shimakura:ss98}
\bibinfo{author}{\bibfnamefont{T.}~\bibnamefont{Shimakura}},
  \bibinfo{author}{\bibfnamefont{H.}~\bibnamefont{Minoda}},
  \bibinfo{author}{\bibfnamefont{Y.}~\bibnamefont{Tanishiro}},
  \bibnamefont{and} \bibinfo{author}{\bibfnamefont{K.}~\bibnamefont{Yagi}},
  \bibinfo{journal}{Surf. Sci.} \textbf{\bibinfo{volume}{407}},
  \bibinfo{pages}{L657} (\bibinfo{year}{1998}).

\bibitem[{\citenamefont{G{\"u}rl{\"u} et~al.}(2003)\citenamefont{G{\"u}rl{\"u},
  Adam, Zandvliet, and Poelsema}}]{Gurlu:apl03}
\bibinfo{author}{\bibfnamefont{O.}~\bibnamefont{G{\"u}rl{\"u}}},
  \bibinfo{author}{\bibfnamefont{O.~A.~O.} \bibnamefont{Adam}},
  \bibinfo{author}{\bibfnamefont{H.~J.~W.} \bibnamefont{Zandvliet}},
  \bibnamefont{and} \bibinfo{author}{\bibfnamefont{B.}~\bibnamefont{Poelsema}},
  \bibinfo{journal}{Appl. Phys. Lett.} \textbf{\bibinfo{volume}{83}},
  \bibinfo{pages}{4610} (\bibinfo{year}{2003}).

\bibitem[{\citenamefont{{\"O}ncel et~al.}(2005)\citenamefont{{\"O}ncel, van
  Houselt, Huijben, Hallback, G{\"u}rl{\"u}, Zandvliet, and
  Poelsema}}]{Oncel:prl05}
\bibinfo{author}{\bibfnamefont{N.}~\bibnamefont{{\"O}ncel}},
  \bibinfo{author}{\bibfnamefont{A.}~\bibnamefont{van Houselt}},
  \bibinfo{author}{\bibfnamefont{J.}~\bibnamefont{Huijben}},
  \bibinfo{author}{\bibfnamefont{A.~S.} \bibnamefont{Hallback}},
  \bibinfo{author}{\bibfnamefont{O.}~\bibnamefont{G{\"u}rl{\"u}}},
  \bibinfo{author}{\bibfnamefont{H.~J.~W.} \bibnamefont{Zandvliet}},
  \bibnamefont{and} \bibinfo{author}{\bibfnamefont{B.}~\bibnamefont{Poelsema}},
  \bibinfo{journal}{Phys. Rev. Lett.} \textbf{\bibinfo{volume}{95}},
  \bibinfo{pages}{116801} (\bibinfo{year}{2005}).

\bibitem[{\citenamefont{Sch{\"a}fer et~al.}(2006)\citenamefont{Sch{\"a}fer,
  Schrupp, Preisinger, and Claessen}}]{Schafer:prb06}
\bibinfo{author}{\bibfnamefont{J.}~\bibnamefont{Sch{\"a}fer}},
  \bibinfo{author}{\bibfnamefont{D.}~\bibnamefont{Schrupp}},
  \bibinfo{author}{\bibfnamefont{M.}~\bibnamefont{Preisinger}},
  \bibnamefont{and} \bibinfo{author}{\bibfnamefont{R.}~\bibnamefont{Claessen}},
  \bibinfo{journal}{Phys. Rev. B} \textbf{\bibinfo{volume}{74}},
  \bibinfo{eid}{041404(R)} (\bibinfo{year}{2006}).

\bibitem[{\citenamefont{Smit et~al.}(2001)\citenamefont{Smit, Untiedt, Yanson,
  and van Ruitenbeek}}]{Smit:prl01}
\bibinfo{author}{\bibfnamefont{R.~H.~M.} \bibnamefont{Smit}},
  \bibinfo{author}{\bibfnamefont{C.}~\bibnamefont{Untiedt}},
  \bibinfo{author}{\bibfnamefont{A.~I.} \bibnamefont{Yanson}},
  \bibnamefont{and} \bibinfo{author}{\bibfnamefont{J.~M.} \bibnamefont{van
  Ruitenbeek}}, \bibinfo{journal}{Phys. Rev. Lett.}
  \textbf{\bibinfo{volume}{87}}, \bibinfo{pages}{266102}
  (\bibinfo{year}{2001}).

\bibitem[{\citenamefont{Wang et~al.}(2004)\citenamefont{Wang, Li, and
  Altman}}]{Wang:prb04}
\bibinfo{author}{\bibfnamefont{J.}~\bibnamefont{Wang}},
  \bibinfo{author}{\bibfnamefont{M.}~\bibnamefont{Li}}, \bibnamefont{and}
  \bibinfo{author}{\bibfnamefont{E.~I.} \bibnamefont{Altman}},
  \bibinfo{journal}{Phys. Rev. B} \textbf{\bibinfo{volume}{70}},
  \bibinfo{pages}{233312} (\bibinfo{year}{2004}).

\bibitem[{\citenamefont{Wang et~al.}(2005)\citenamefont{Wang, Li, and
  Altman}}]{Wang:ss05}
\bibinfo{author}{\bibfnamefont{J.}~\bibnamefont{Wang}},
  \bibinfo{author}{\bibfnamefont{M.}~\bibnamefont{Li}}, \bibnamefont{and}
  \bibinfo{author}{\bibfnamefont{E.~I.} \bibnamefont{Altman}},
  \bibinfo{journal}{Surf. Sci.} \textbf{\bibinfo{volume}{596}},
  \bibinfo{pages}{126} (\bibinfo{year}{2005}).

\bibitem[{\citenamefont{Vanpoucke and Brocks}(2008)}]{vanpoucke:prb2008R}
\bibinfo{author}{\bibfnamefont{D.~E.~P.} \bibnamefont{Vanpoucke}}
  \bibnamefont{and} \bibinfo{author}{\bibfnamefont{G.}~\bibnamefont{Brocks}},
  \bibinfo{journal}{Phys. Rev. B} \textbf{\bibinfo{volume}{77}},
  \bibinfo{eid}{241308(R)} (\bibinfo{year}{2008}).

\bibitem[{\citenamefont{Stekolnikov et~al.}(2008)\citenamefont{Stekolnikov,
  Bechstedt, Wisniewski, Schafer, and Claessen}}]{stekolnikov:prl08}
\bibinfo{author}{\bibfnamefont{A.~A.} \bibnamefont{Stekolnikov}},
  \bibinfo{author}{\bibfnamefont{F.}~\bibnamefont{Bechstedt}},
  \bibinfo{author}{\bibfnamefont{M.}~\bibnamefont{Wisniewski}},
  \bibinfo{author}{\bibfnamefont{J.}~\bibnamefont{Schafer}}, \bibnamefont{and}
  \bibinfo{author}{\bibfnamefont{R.}~\bibnamefont{Claessen}},
  \bibinfo{journal}{Phys. Rev. Lett.} \textbf{\bibinfo{volume}{100}},
  \bibinfo{eid}{196101} (\bibinfo{year}{2008}).

\bibitem[{\citenamefont{Bl{\"o}chl}(1994)}]{Blochl:prb94}
\bibinfo{author}{\bibfnamefont{P.~E.} \bibnamefont{Bl{\"o}chl}},
  \bibinfo{journal}{Phys. Rev. B} \textbf{\bibinfo{volume}{50}},
  \bibinfo{pages}{17953} (\bibinfo{year}{1994}).

\bibitem[{\citenamefont{Kresse and Joubert}(1999)}]{Kresse:prb99}
\bibinfo{author}{\bibfnamefont{G.}~\bibnamefont{Kresse}} \bibnamefont{and}
  \bibinfo{author}{\bibfnamefont{D.}~\bibnamefont{Joubert}},
  \bibinfo{journal}{Phys. Rev. B} \textbf{\bibinfo{volume}{59}},
  \bibinfo{pages}{1758} (\bibinfo{year}{1999}).

\bibitem[{\citenamefont{Ceperley and Alder}(1980)}]{CA:prl1980}
\bibinfo{author}{\bibfnamefont{D.~M.} \bibnamefont{Ceperley}} \bibnamefont{and}
  \bibinfo{author}{\bibfnamefont{B.~J.} \bibnamefont{Alder}},
  \bibinfo{journal}{Phys. Rev. Lett.} \textbf{\bibinfo{volume}{45}},
  \bibinfo{pages}{566} (\bibinfo{year}{1980}).

\bibitem[{\citenamefont{Kresse and Hafner}(1993)}]{Kresse:prb93}
\bibinfo{author}{\bibfnamefont{G.}~\bibnamefont{Kresse}} \bibnamefont{and}
  \bibinfo{author}{\bibfnamefont{J.}~\bibnamefont{Hafner}},
  \bibinfo{journal}{Phys. Rev. B} \textbf{\bibinfo{volume}{47}},
  \bibinfo{pages}{558} (\bibinfo{year}{1993}).

\bibitem[{\citenamefont{Kresse and Furthm\"uller}(1996)}]{Kresse:prb96}
\bibinfo{author}{\bibfnamefont{G.}~\bibnamefont{Kresse}} \bibnamefont{and}
  \bibinfo{author}{\bibfnamefont{J.}~\bibnamefont{Furthm\"uller}},
  \bibinfo{journal}{Phys. Rev. B} \textbf{\bibinfo{volume}{54}},
  \bibinfo{pages}{11169} (\bibinfo{year}{1996}).

\bibitem[{\citenamefont{Monkhorst and Pack}(1976)}]{Monkhorst:prb76}
\bibinfo{author}{\bibfnamefont{H.~J.} \bibnamefont{Monkhorst}}
  \bibnamefont{and} \bibinfo{author}{\bibfnamefont{J.~D.} \bibnamefont{Pack}},
  \bibinfo{journal}{Phys. Rev. B} \textbf{\bibinfo{volume}{13}},
  \bibinfo{pages}{5188} (\bibinfo{year}{1976}).

\bibitem[{\citenamefont{Needels et~al.}(1987)\citenamefont{Needels, Payne, and
  Joannopoulos}}]{Needels:prl87}
\bibinfo{author}{\bibfnamefont{M.}~\bibnamefont{Needels}},
  \bibinfo{author}{\bibfnamefont{M.~C.} \bibnamefont{Payne}}, \bibnamefont{and}
  \bibinfo{author}{\bibfnamefont{J.~D.} \bibnamefont{Joannopoulos}},
  \bibinfo{journal}{Phys. Rev. Lett.} \textbf{\bibinfo{volume}{58}},
  \bibinfo{pages}{1765} (\bibinfo{year}{1987}).

\bibitem[{\citenamefont{Needels et~al.}(1988)\citenamefont{Needels, Payne, and
  Joannopoulos}}]{Needels:prb88}
\bibinfo{author}{\bibfnamefont{M.}~\bibnamefont{Needels}},
  \bibinfo{author}{\bibfnamefont{M.~C.} \bibnamefont{Payne}}, \bibnamefont{and}
  \bibinfo{author}{\bibfnamefont{J.~D.} \bibnamefont{Joannopoulos}},
  \bibinfo{journal}{Phys. Rev. B} \textbf{\bibinfo{volume}{38}},
  \bibinfo{pages}{5543} (\bibinfo{year}{1988}).

\bibitem[{\citenamefont{Jenkins and Srivastava}(1996)}]{Jenkins:jp96}
\bibinfo{author}{\bibfnamefont{S.~J.} \bibnamefont{Jenkins}} \bibnamefont{and}
  \bibinfo{author}{\bibfnamefont{G.~P.} \bibnamefont{Srivastava}},
  \bibinfo{journal}{J. Phys.: Condens. Matter.} \textbf{\bibinfo{volume}{8}},
  \bibinfo{pages}{6641} (\bibinfo{year}{1996}).

\bibitem[{\citenamefont{Gay and Srivastava}(1999)}]{Gay:prb99}
\bibinfo{author}{\bibfnamefont{S.~C.~A.} \bibnamefont{Gay}} \bibnamefont{and}
  \bibinfo{author}{\bibfnamefont{G.~P.} \bibnamefont{Srivastava}},
  \bibinfo{journal}{Phys. Rev. B} \textbf{\bibinfo{volume}{60}},
  \bibinfo{pages}{1488} (\bibinfo{year}{1999}).

\bibitem[{fn:({\natexlab{a}})}]{fn:dimernomen}
\bibinfo{note}{For the sake of simplicity and ease of writing we refer to any
  bound atom-pair on the surface and in its top-layer as a dimer, even if it
  consists of two different atoms. If the atom-types are important we will
  refer to the dimer as an X-Y dimer, with X and Y its constituent atoms. Also
  a shorter notation will be used where Ge dimers is used to refer to Ge-Ge
  dimers and Pt dimers is used to refer to dimers containing at least $1$ Pt
  atom. For the latter it will follow from the context if this refers to Pt-Ge
  dimers or Pt-Pt dimers.}

\bibitem[{fn:({\natexlab{b}})}]{fn:udcomm}
\bibinfo{note}{Note that the $u$ and $d$ index only indicate the position in
  the Ge(001) reconstruction geometry, and tells nothing about the relative
  $z$-position of the Pt atoms with regard to the Ge atom in the specific Pt-Ge
  dimer.}

\bibitem[{fn:({\natexlab{c}})}]{fn:altformenerg}
\bibinfo{note}{An alternative way of calculating the formation energy (which
  gives exactly the same results) is by calculating the surface formation
  energy $E_{sb}$ of the $\beta$-surface using the expression, \[E_{sb} =
  E_T[\beta_{\mathrm{x}}]-N_{Pt}E_{Pt}-N_{Ge}E_{Ge}, \] where
  $E_T[\beta_{\mathrm{x}}]$ is the total energy of the
  $\beta_{\mathrm{x}}$-surface, $N_{X}$ the number of atoms of type $X$ in the
  slab with $E_{X}$ the corresponding bulk energy per atom. This formation
  energy we can now consider to be the sum of the formation energies of the
  separate surface dimers. Because the surface consists of two types of dimers,
  it is reasonable to assume that they will contribute differently to the total
  formation energy. The formation energy of a Ge-Ge dimer is found by
  calculating the formation energy per surface dimer for a Ge(001)
  $b(2\times1)$ surface reconstruction. Substraction of the formation energy
  caused by the Ge-Ge dimers gives: \[E_{sb2}=E_{sb}-N_{Ge-dimers}E_{s\
  Ge-dimer},\] with $N_{Ge-dimers}$ the number of Ge-Ge dimers in the
  $\beta$-surface and $E_{s\ Ge-dimer}$ the formation energy of a Ge dimer in
  the $b(2\times1)$ surface reconstruction. The resulting energy $E_{sb2}$ can
  then be considered the contribution by the Pt-Ge dimers to $E_{sb}$.\\ If we
  again subtract the formation energy $E_{s\ Ge-dimer}$ from the calculated
  energy contribution pet Pt-Ge dimer, $E_{sb2}$/$N_{Pt-dimers}$, the formation
  energy per Pt-Ge dimer as given in Eq.~\eqref{eq:1_Formation_energy} is
  obtained.\\ \indent In this derivation the physical meaning of the formation
  energy becomes clear: ``the energy gained/needed to replace a Ge-Ge dimer by
  a Pt containing dimer''. This is true under the assumption that there is no
  contribution to the surface energy due to dimer-dimer interactions or (as we
  will do) if all changes herein are attributed to the Pt containing dimers and
  they are assumed to be equal in a single geometry. This allows us to estimate
  the energy contributions to the surface formation energy due to inter- and
  intra dimer row interactions, by comparison of similar geometries.}

\bibitem[{\citenamefont{Niranjan et~al.}(2007)\citenamefont{Niranjan, Kleinman,
  and Demkov}}]{Niranjan:prb07}
\bibinfo{author}{\bibfnamefont{M.~K.} \bibnamefont{Niranjan}},
  \bibinfo{author}{\bibfnamefont{L.}~\bibnamefont{Kleinman}}, \bibnamefont{and}
  \bibinfo{author}{\bibfnamefont{A.~A.} \bibnamefont{Demkov}},
  \bibinfo{journal}{Phys. Rev. B} \textbf{\bibinfo{volume}{75}},
  \bibinfo{pages}{085326} (\bibinfo{year}{2007}).

\bibitem[{\citenamefont{Tersoff and Hamann}(1985)}]{Tersoff:prb85}
\bibinfo{author}{\bibfnamefont{J.}~\bibnamefont{Tersoff}} \bibnamefont{and}
  \bibinfo{author}{\bibfnamefont{D.~R.} \bibnamefont{Hamann}},
  \bibinfo{journal}{Phys. Rev. B} \textbf{\bibinfo{volume}{31}},
  \bibinfo{pages}{805} (\bibinfo{year}{1985}).

\bibitem[{\citenamefont{G{\"u}rl{\"u}}(2004)}]{Gurlu:thesis}
\bibinfo{author}{\bibfnamefont{O.}~\bibnamefont{G{\"u}rl{\"u}}}, Ph.D. thesis,
  \bibinfo{school}{University of Twente} (\bibinfo{year}{2004}).

\bibitem[{\citenamefont{van Houselt et~al.}(2006)\citenamefont{van Houselt,
  {\"O}ncel, Poelsema, and Zandvliet}}]{Houselt:nanol06}
\bibinfo{author}{\bibfnamefont{A.}~\bibnamefont{van Houselt}},
  \bibinfo{author}{\bibfnamefont{N.}~\bibnamefont{{\"O}ncel}},
  \bibinfo{author}{\bibfnamefont{B.}~\bibnamefont{Poelsema}}, \bibnamefont{and}
  \bibinfo{author}{\bibfnamefont{H.}~\bibnamefont{Zandvliet}},
  \bibinfo{journal}{Nano Letters} \textbf{\bibinfo{volume}{6}},
  \bibinfo{pages}{1439} (\bibinfo{year}{2006}).

\bibitem[{fn:({\natexlab{d}})}]{fn:Ge_BG}
\bibinfo{note}{The zero, or nearly zero, band gap for germanium is a known
  failure of LDA. Although a band gap can be opened using for example the GW
  method, it is not necessary for the qualitative comparison we are making
  here.}

\bibitem[{\citenamefont{G{\"u}rl{\"u} et~al.}(2004)\citenamefont{G{\"u}rl{\"u},
  Zandvliet, Poelsema, Dag, and Ciraci}}]{Gurlu:prb04}
\bibinfo{author}{\bibfnamefont{O.}~\bibnamefont{G{\"u}rl{\"u}}},
  \bibinfo{author}{\bibfnamefont{H.~J.~W.} \bibnamefont{Zandvliet}},
  \bibinfo{author}{\bibfnamefont{B.}~\bibnamefont{Poelsema}},
  \bibinfo{author}{\bibfnamefont{S.}~\bibnamefont{Dag}}, \bibnamefont{and}
  \bibinfo{author}{\bibfnamefont{S.}~\bibnamefont{Ciraci}},
  \bibinfo{journal}{Phys. Rev. B} \textbf{\bibinfo{volume}{70}},
  \bibinfo{pages}{085312} (\bibinfo{year}{2004}).

\bibitem[{\citenamefont{Zeng and Elsayed-Ali}(2002)}]{Zeng:ss02}
\bibinfo{author}{\bibfnamefont{X.}~\bibnamefont{Zeng}} \bibnamefont{and}
  \bibinfo{author}{\bibfnamefont{H.~E.} \bibnamefont{Elsayed-Ali}},
  \bibinfo{journal}{Surf. Sci.} \textbf{\bibinfo{volume}{497}},
  \bibinfo{pages}{373} (\bibinfo{year}{2002}).

\end{thebibliography}

\end{document}